\documentclass[useAMS]{mn2e}

\usepackage{graphicx}	
\usepackage{amsmath}	
\usepackage{amssymb}	
\usepackage{multicol}       
\usepackage{bm}		
\usepackage{pdflscape}	
\usepackage[utf8]{inputenc}
\usepackage[T1]{fontenc}
\usepackage{ae,aecompl}

\title[{\it GAIA} DR3 wide binary kinematics] {Internal kinematics of {\it GAIA} DR3 wide binaries: anomalous behaviour in the
  low acceleration regime.} 

\author[X. Hernandez] {X. Hernandez$^{1}$\\ 
$^{1}$Instituto de Astronom\'{\i}a, Universidad Nacional Aut\'{o}noma de M\'{e}xico,
  Apartado Postal 70--264 C.P. 04510 M\'exico D.F. M\'exico. \\
}

\date{Released 15/04/2022}

\begin{document}

\label{firstpage}

\maketitle

\begin{abstract}
  The {\it Gaia} eDR3 catalogue has recently been used to study statistically the internal kinematics of wide binary populations
  using relative velocities of the two component stars, $\Delta V$, total binary masses, $m_{B}$, and separations, $s$. For $s \gtrsim 0.01$ pc, these
  binaries probe the low acceleration $a \lesssim 2a_{0}$ regime where gravitational anomalies usually attributed to dark matter are observed
  in the flat
  rotation curves of spiral galaxies, where $a_{0}\approx 1.2\times 10^{-10}$m s$^{-2}$ is the acceleration scale of MOND. Such experiments test the degree of
  generality of these anomalies, by exploring the same acceleration regime using independent astronomical systems of vastly smaller mass and size.
  A signal above Newtonian expectations has been observed when $a \lesssim 2a_{0}$, alternatively interpreted as evidence of a modification of gravity,
  or as due to kinematic contaminants; undetected stellar components, unbound
  encounters or spurious projection effects. Here I take advantage of the enhanced DR3 {\it Gaia} catalogue to perform a more rigorous 
  study of the internal kinematics of wide binaries than what has previously been possible. Internally determined {\it Gaia} stellar
  masses and estimates of binary probabilities for each star using spectroscopic information, together with a larger sample of radial velocities, allow
  for a significant improvement in the analysis and careful exclusion of possible kinematic contaminants. Resulting $\Delta V$ scalings
  accurately tracing Newtonian expectations for the high acceleration regime, but markedly inconsistent with these expectations in the low acceleration
  one, are obtained. A non-Newtonian low acceleration phenomenology is thus confirmed.

\end{abstract}


\section{Introduction}

The presence of gravitational anomalies in the low acceleration regime of $a<a_{0}\approx 1.2 \times 10^{-10}$ m s$^{-2}$ at galactic scales has been alternatively
interpreted as evidence for a dominant dark matter component of unknown origin and as yet lacking any direct confirmation, or as indicating a change of regime 
in the structure of gravity, generically termed Modified Gravity (e.g. Milgrom 1983), or even a validity limit for Newton's second law, Modified Inertia proposals,
e.g. Milgrom (1994), Milgrom (2022). Empirically, to first order such gravitationally anomalous regime is characterised by flat rotation curves at an amplitude
satisfying the baryonic Tully-Fisher relation, $V_{TF}=(GMa_{0})^{1/4}=0.35(M/M_{\odot})^{1/4}$ km s$^{-1}$, where $M$ refers to the total baryonic mass of a galaxy,
McGaugh et al. (2000), {  Lelli et al. (2017)}.

Deciding between the alternative interpretations could benefit from exploring the low acceleration regime in different astronomical contexts, to obtain evidence
as to the generality, or lack thereof, of the gravitational anomalies present in the outskirts of spiral galaxies. Steps in this direction have been taken by
studies probing the possible presence of Tully-Fisher phenomenology in pressure supported galactic systems by e.g. Jimenez et al. (2013), Durazo et al. (2018)
and Chae et al. (2020a), who find asymptotic velocity dispersion amplitudes also scaling with the fourth root of total baryonic masses. Extensions towards
dwarf galaxies have also been explored, e.g. McGaugh et al. (2021). In going to smaller pressure supported systems, e.g. Scarpa et al. (2003), Hernandez et al.
(2012b) and Hernandez \& Lara-D I (2020) show asymptotically flat velocity dispersion profiles of Galactic Globular Clusters also presenting the same empirical
scalings with total baryonic mass of the baryonic Tully-Fisher relation, albeit the results of Claydon et al. (2017), who present an explanation for the
observed Globular Cluster phenomenology within a standard gravitational framework.

As first identified in Hernandez et al. (2012a), large samples of wide binaries with internal separations larger than about $0.01$ pc offer a probe of the low
acceleration $a \lesssim 2a_{0}$ regime at mass and length scales many orders of magnitude below those of spiral galaxies, and even below the ones mentioned above,
where any dark matter contribution would be negligible. {  Even at the widest separations considered here, of 0.06 pc, the local dark matter density inferred
under a Newtonian framework of $0.01 M_{\odot}$ pc$^{-3}$ (e.g. Read 2014), would only imply about $1\times 10^{-5} M_{\odot}$ of dark matter within the wide binary
orbit, and hence only a contribution of order one part in $10^{5}$ of the total mass of the system, with individual stellar masses not far from $0.8 M_{\odot}$.}

The internal kinematics of wide binary samples as a test of gravity in the low acceleration regime have more recently been considered by: Scarpa et al. (2017),
Banik \& Zhao (2018), Pittordis \& Sutherland (2018), Hernandez et al. (2019a), Banik (2019), Pittordis \& Sutherland (2019), Acedo (2020), Hernandez et al.
(2022), Manchanda et al. (2023) and Pittordis \& Sutherland (2023). The presence of anomalous internal velocities in wide binaries is now well established,
with various interpretations having been proposed. In Hernandez et al. (2022) we used {\it Gaia} eDR3 to show that on reaching the low acceleration regime of
sufficiently separated binaries, a qualitative regime change appears where the binned rms projected relative velocity for wide binary populations ceases to
drop along Keplerian expectations, and settles to values consistent with the baryonic Tully-Fisher scaling of spiral galaxies, to argue in favour of a change
in regime for the physics describing the problem.

On the other hand, Clarke (2020) simulating wide binary populations and Pittordis et al. (2023) using {\it Gaia} eDR3 data, show that introducing a
hypothetical population of hidden tertiaries, cases where one or both components of an observed binary harbour an undetected stellar companion, results in a
kinematic contaminant which if chosen judiciously can explain the observations within a Newtonian framework. This population of hidden tertiaries hence becomes
a prediction of Newtonian gravity, which fortunately has recently been shown to lie within the reach of independent confirmation through dedicated follow-up
studies using a variety of readily available techniques, Manchanda et al. (2023).

In this paper I explore from an empirical approach, as an extension of our previous study of Hernandez et al. (2022), the kinematics of Solar Neighbourhood
wide binaries and the mass-velocity scalings these present, taking advantage for the first time in this context of the recent {\it Gaia} DR3 {  catalogue}.
This latest data release benefits from direct mass estimates using spectroscopic information and the {\it Gaia} work package FLAME for a sub-set of stellar
sources, an internally assessed binary probability for each star, the {\bf CLASSPROB\_DSC\_COMBMOD\_BINARYSTAR} parameter, an inference using spectral,
photometric and astrometric information, henceforth $B_{P}$, as well about $4.7 \times$ more stars containing velocities along the line of sight. All of the
above improvements on {\it Gaia} eDR3 permit a more accurate probing of the problem, with an enlarged sample of stars having radial velocities, useful to
exclude unbound flyby events, and more accurate elimination of kinematic contaminants, e.g. by imposing cuts in the $B_{P}$ parameter of a sample, as well
as allowing for an accurate calibration of the luminosity-mass scalings used in all the studies mentioned above to infer individual stellar masses.

The structure of the paper is as follows: Section 2 describes the sample selection, driven by the philosophy of defining a small but very high quality sample
where the consistency or otherwise of wide binary internal kinematics with Newtonian expectations might be assessed. Section 3 shows the calibration of a
simple mass-luminosity relation to the high quality spectroscopically determined {\it Gaia} DR3 individual masses, which is then used throughout. Section 4
presents results for two samples having different distance cuts so as to obtain a measure of distance dependent effects. Section 5 gives a comparison to
recent independent work, and section 6 a final discussion. {  Additional consistency checks appear in the appendix.}

\section{Sample Selection}

 \begin{figure}
 \vskip 0pt
 \hskip -5pt
 \includegraphics[height=7.0cm,width=8.5cm]{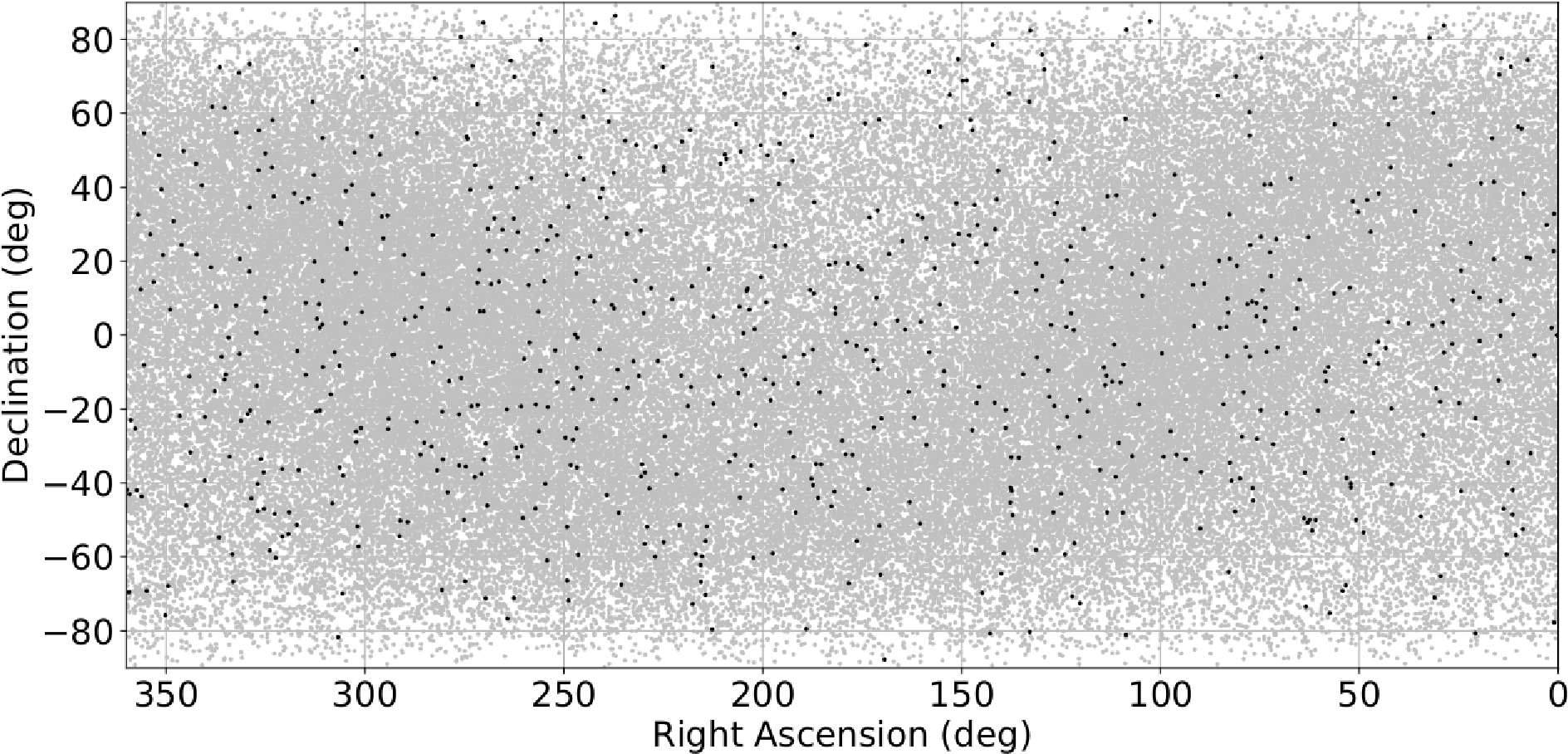}
 \caption{{  The grey dots show a} sky plot of 95,899 binary pairs with $(D/$pc$)<333$, $(S/N)_{\varpi}>100$ and $(S/N)_{G}>5$ in each star, where all binary
   candidates having stars in common have already been discarded. The 1352 binary pairs within $(D/$pc$)<125$ and further satisfying $RUWE<1.2$, $B_{p}<0.2$,
   $\Delta V_{LOS}<4 km/s$ and $R_{P}$,$G$,$B_{P}$ and proper motion signal-to-noise values $>20$ are shown as black dots. Notice no conspicuous groupings or
   local clusters remain after the aggressive de-grouping procedure used.}
 \end{figure}

 \begin{figure*}
      \includegraphics[height=7.0cm,width=8.8cm]{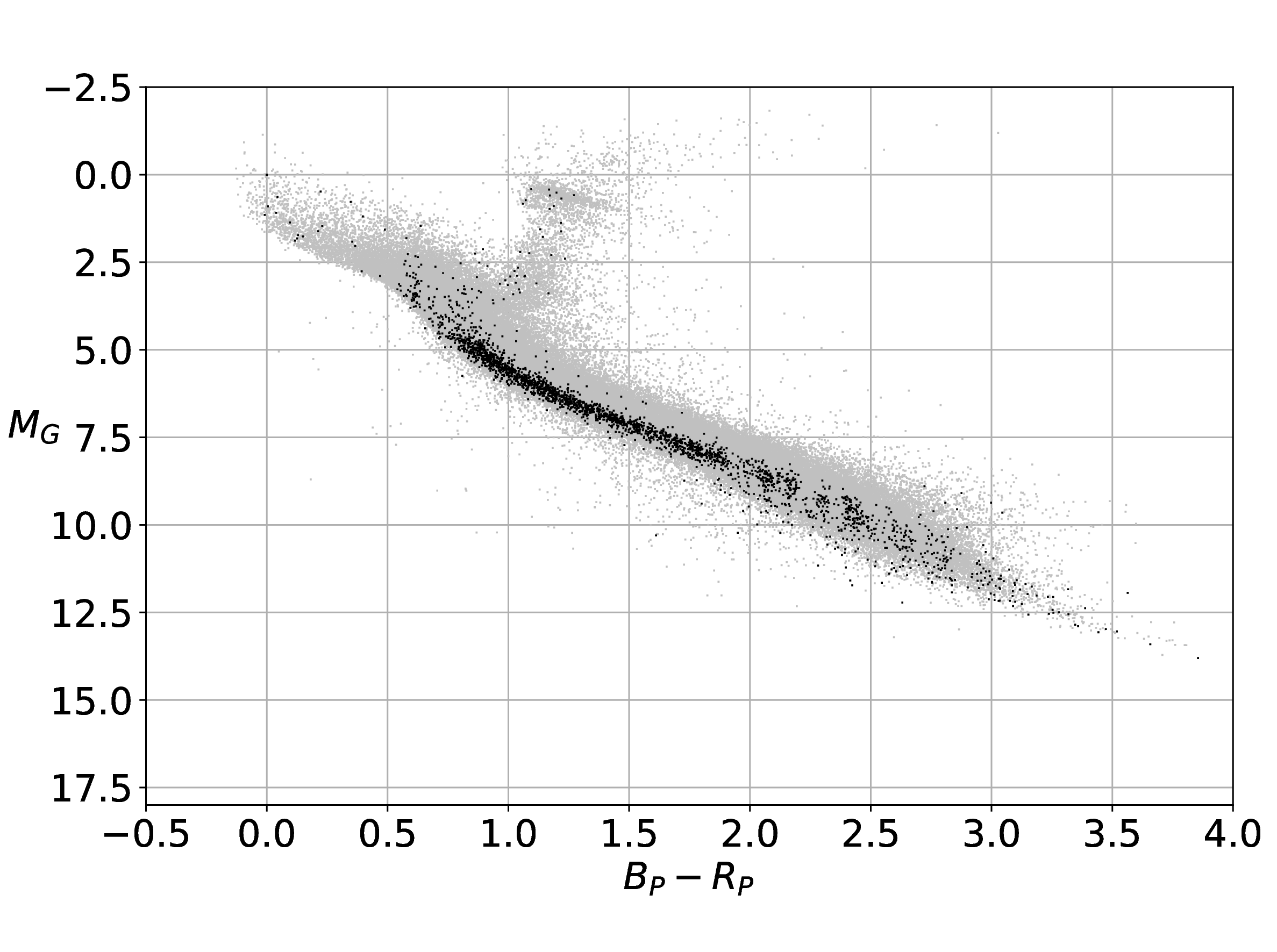}
     \hspace*{-5pt}
      \includegraphics[height=7.0cm,width=8.8cm]{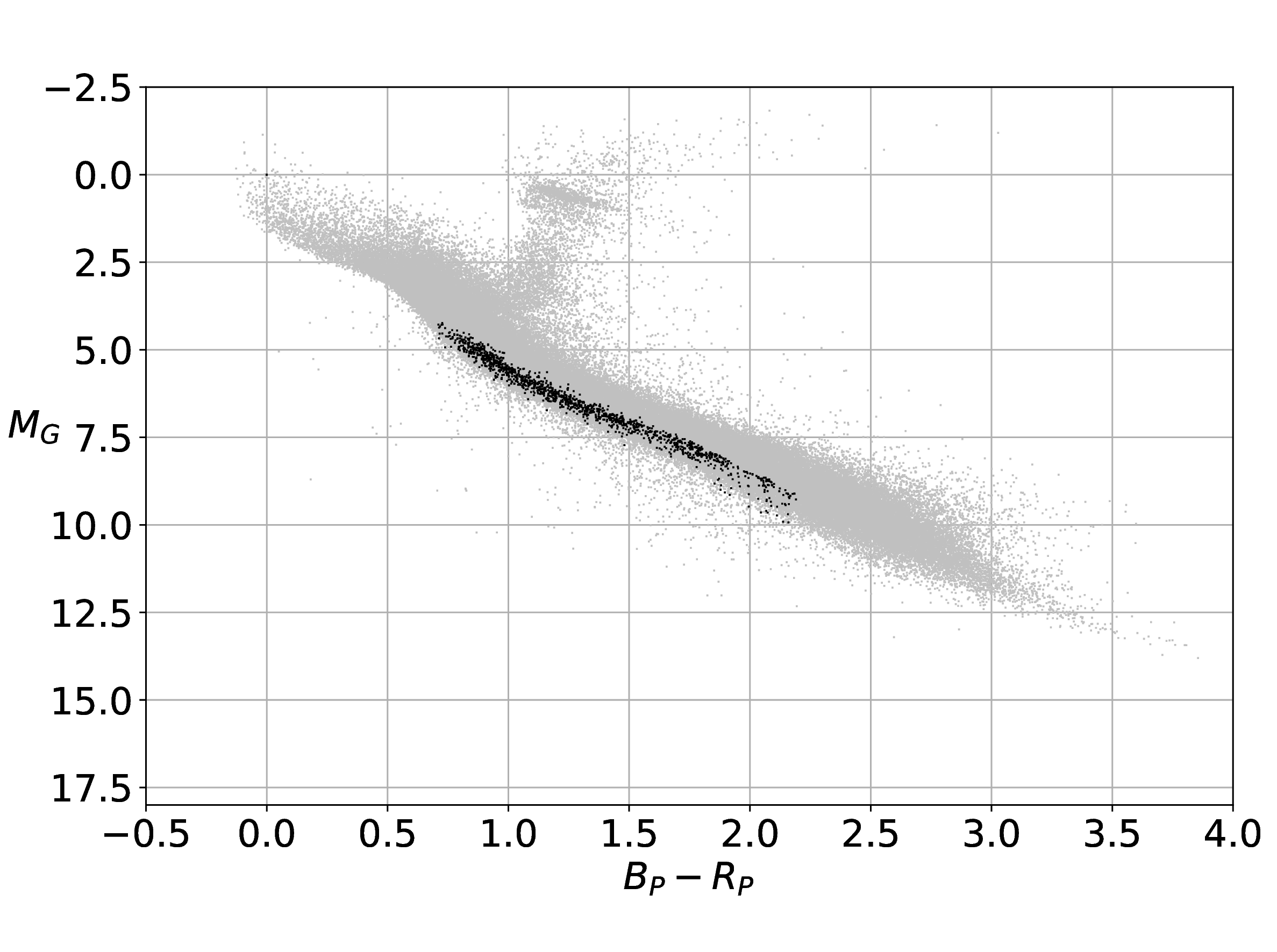}
      \caption{{  Left(a)}: CMD for the stars in the 95,899 binaries mentioned in Fig. 1, shown as grey dots. The stars for the 1,352 binary pairs shown as
        black dots in Fig. 1 appear here as black dots. {  Right(b)}: CMD for the stars in the 688 binary pairs within the colour magnitude region selected
        to eliminate photometric binaries as member stars of final selected binaries, and minimise the inclusion of hidden tertiaries in the kinematic samples.}   
 \end{figure*}

The sample selection used is a modified version of the {\it Gaia} one used by El-Badry \& Rix (2018) who present and test a wide binary catalogue with a distance
cut of $D<200$pc. Accounting for projection effects {  and using} simulations modelling reasonable distributions of ellipticities and undetected companions to
estimate completion factors, these authors report a level of contamination of below $0.2 \%$. Based on this same approach Tian et al. (2020) produced a lower
quality but more extensive catalogue with a distance cut of $D<4$kpc containing 800 000 binary candidates.

I begin with a {\it Gaia} search returning all stars within $333$pc with accurate parallaxes and G magnitudes having a signal-to-noise ratio of
$(S/N)_{\varpi}>100$ and {a signal-to-noise ratio in the {\it Gaia} G band of}
$(S/N)_{G}>5$. I then search within a projected $0.5$pc radius on the plane of the sky about each star for a potential binary companion. Any
resulting pair is then accepted as a binary candidate, provided the difference in distance along the line of sight between both stars, $\Delta D$, is smaller than
twice the projected separation between both stars, $2s$, at a $3 \sigma$ level, i.e., {  that $\Delta D-3\sigma_{\Delta D} < 2s < \Delta D+3\sigma_{\Delta D} $.}
A 192 HEALPix scheme is used, to limit the number of lost binaries where component stars lie in adjacent HEALPix.

As noticed in El-Badry \& Rix (2018), the fixed {\it Gaia} resolution implies that
as the distance cut of the sample increases, a growing fraction of close binaries will be lost. Hence, the sample will have a distance-dependent completion factor.
This is not a concern, as the objective is not to ensure valid candidates are all included with a fixed probability, but that unsound binary candidates are excluded.
I modify the original El-Badry \& Rix (2018) selection criteria, to remove the condition that the relative velocity between the two components of a binary system should
be within Newtonian expectations, as it is precisely the validity of this assumption that is being probed.

The next step is to aggressively eliminate any binary candidates which might be under the gravitational influence of a third star, or indeed be part of a tertiary
system. I search for any stars which are members of more than one candidate binary, and remove all such binary candidates. Thus, I do not try to decide which
binary a given star flagged as a member of two or more binary candidate systems belongs to, but eliminate all candidate binaries which contain individual stars forming
part of more than one such system. Given the parallax signal-to-noise ratio of the sample, for a binary system at an average distance for the final binaries used
of $ \approx 100$pc, by construction, no other {\it Gaia} sources remain within $1$pc along the line of sight at the $1\sigma$ level. For binary systems in the
critical range of internal separations of $\approx 0.03$pc (see section 4), this implies an isolation factor along the line of sight of about $ 30\times$ the
internal binary separation. On the plane of the sky, the initial search criteria ensures an isolation factor of $5/3\times$ larger than along the line of sight
at a $1\sigma$ level, ensuring the final selected binaries are indeed free from kinematic contamination from other {\it Gaia} sources, to a very large degree.
Of the close to 10 million binary candidates originally identified, the above de-grouping algorithm leaves only 97,251 binary pairs. 

Next, I apply a series of data quality cuts to each of the stars of each candidate binary system, and remove any such system where either of the constituent stars
fails the cut. First, $R_{P}$, $G$, and $B_{P}$ magnitude signal-to-noise cuts of $>20$, which given the strong correlations between $(S/N)_{\varpi}$ and magnitude
signal-to-noise levels, actually {  excludes} very few binary candidates, but ensures no suspicious sources are being considered. As discussed in the introduction,
a primary concern is the possible kinematic contamination from hidden tertiaries, which is addressed through a variety of quality cuts. The first is the use of the
{\it Gaia} internal binary probability field, $B_{P}$, which provides a first order estimate of the probability that each individual {\it Gaia} source is actually
a binary star, which would then make any of our binary candidate systems a hidden tertiary. I impose a $B_{P}<0.2$ quality cut. Also, it has been shown, e.g.
Belokurov et al. (2022), that the probability of a {\it Gaia} source being an unresolved binary is a strong function of the $RUWE$ quality index of the source,
and actually identify a threshold of $RUWE<1.4$ at distances of 1kpc below which hidden tertiaries can be reliably excluded using {\it Gaia} DR2 data. Since the DR3
data used reflect a 34 month timeline, rather than the 22 month one of DR2, and since our 200 pc distance cut-off is much smaller than the 1 kpc which Belokurov
et al. (2022) consider, imposing a stringent $RUWE<1.2$ limit will guarantee a sample relatively free of hidden tertiaries. Again, the two
quality cuts mentioned above are applied to all stars, with any candidate binary system {  being excluded if either of its stars fails any of the cuts.}

 \begin{figure*}
    \includegraphics[height=7.0cm,width=8.8cm]{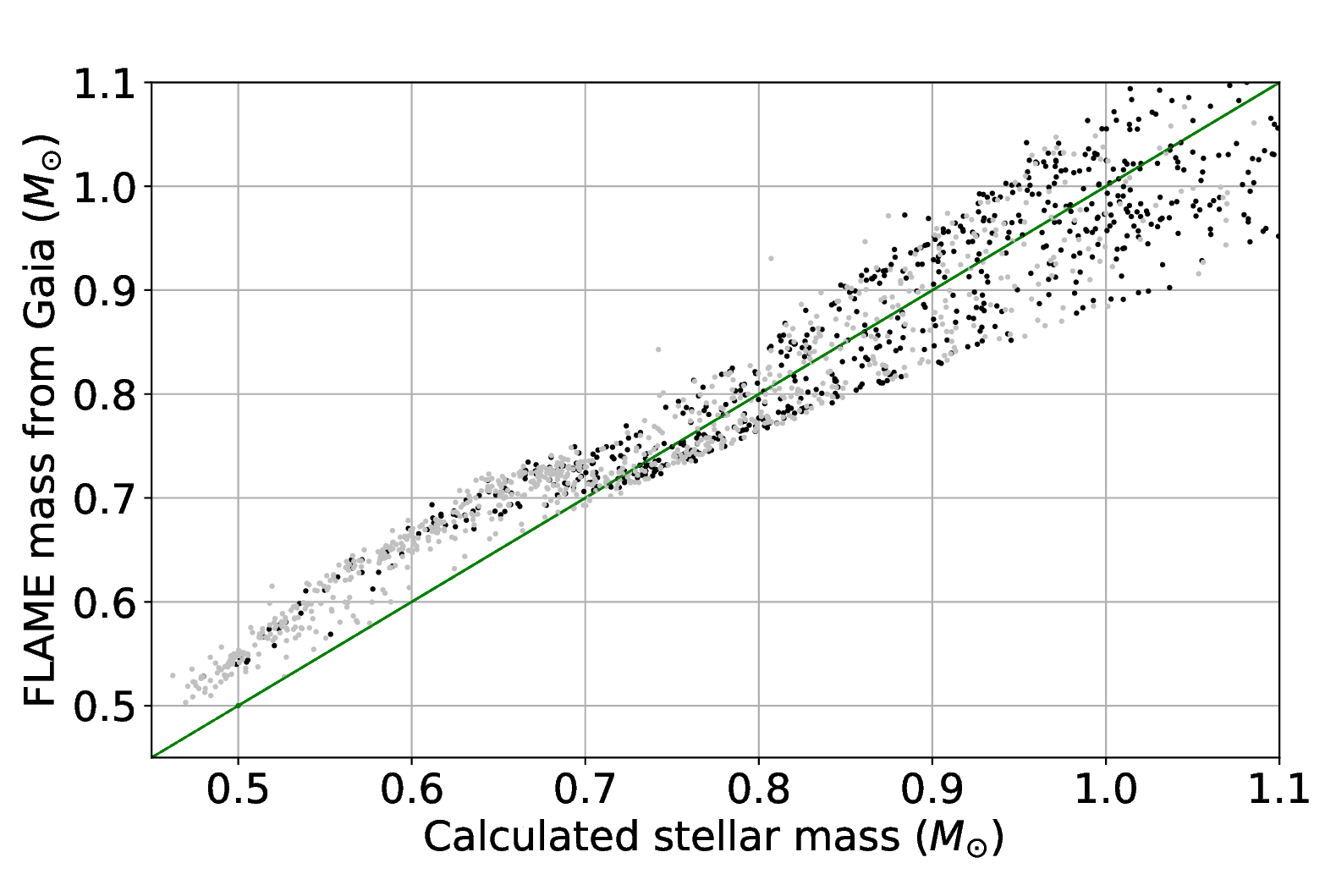}
     \hspace*{-5pt}
    \includegraphics[height=7.0cm,width=8.8cm]{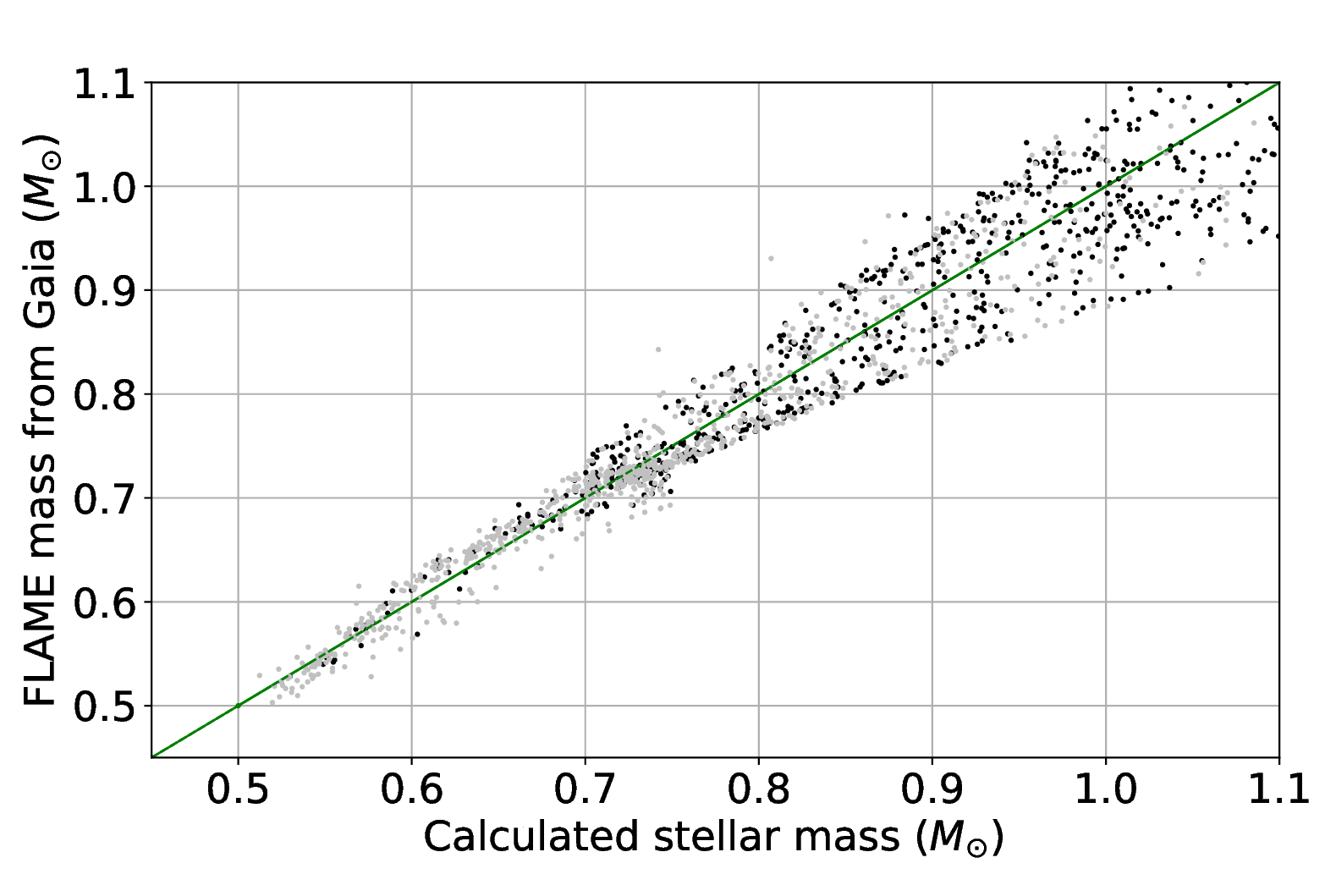}
    \caption{{  Left(a)}:Inferred stellar masses using eq.(1) compared to internally provided spectroscopic {\it Gaia} DR3 masses. A small systematic offset
      is clearly apparent in the low mass range.
      {  Right(b)}: Inferred stellar masses using eq.(1) compared to internally provided spectroscopic {\it Gaia} DR3 {  masses}, after a $0.05 M_{\odot}$ correction
      factor was added to the result of eq.(1), for masses obtained initially below $0.7 M_{\odot}$. No substantial systematic offset remains at this point in the
      mass range shown, where the stars comprising the binaries in the kinematic samples used are found. In both panels grey dots indicate cases where the star in
      question is part of a binary where the other star has no {\it Gaia} DR3 mass reported, while the black dots indicate cases where the star shown is part of a binary
      where the other star also has a reported {\it Gaia} DR3 mass.}   
 \end{figure*}

 Then, I require all stars to have a reported radial velocity measurement in the DR3 catalogue. This cut serves three purposes:
\begin{enumerate}

\begin{item} {As the probability of having a radial velocity measurement in the {\it Gaia} catalogue drops rapidly as the single
star solution degrades, requiring radial velocities serves as a further quality control against hidden tertiaries.}
\end{item}

\begin{item}
Having radial velocities allows to consider full spherical geometric corrections ({  and perspective effects}) when deriving the
relative velocity between both components, Smart (1968). This last correction is only relevant to a few very close and very wide binaries,
indeed, El-Badry (2019) shows that ignoring this correction will result in spuriously elevated relative velocities for wide binaries,
but only for internal separations above about $0.1$pc. {  For my sample such effects are very minor, but are included for completion.}
\end{item}

\begin{item}
Radial velocities allow a filter on projection and unbound flyby systems, which we cut by requiring all our final binary candidate
systems have individual stars showing a radial velocity difference below $4$ km s$^{-1}$.
\end{item}

\end{enumerate}

The battery of quality and hidden tertiary cuts described above culls the original 97,251 binary candidates down to a small and highly curated sample, which for
a distance cut of $D<125$ pc leaves only 1,352 binary pairs. These are shown in a sky plot in Fig. 1 as black dots, where grey dots show the original 97,251
binary candidates. The thorough de-groping strategy described already removes all known nearby clusters and groups from the original binary candidate list,
the strict series of exclusion cuts that then follow leave a small sample of highly isolated wide binaries not showing any evidence of unwanted groups or
clusters.

{  As can be seen in Fig. 1, explicitly excluding the
low Galactic latitudes is unnecessary given the aggressive de-grouping
strategy adopted, as is the specific removal of local known groups. Indeed, the de-grouping implemented
leaves no local over-densities which might correlate with either known local groups
or with the Galactic plane. The use of {\it Gaia} data restricted to distances below 125
pc with very high quality parallaxes makes this cut unnecessary, within that
distance the data show no overcrowding. For the high quality $D<125$ pc sample
from which conclusions are drawn, the final mean signal-to-noise value for the
parallaxes of the stars {  is} 855.4. Even at the
outer limit of this sample, this implies a distance uncertainty of only 0.15 pc,
which ensures crowding against background sources is not an issue. A second
safeguard against the inclusion of spurious binary pairs in overcrowded regions
comes from the relative radial velocity cut introduced, which efficiently
eliminates most projection effects. Indeed, Fig. 1 was included as a check
of this point, the sky plot of binaries remaining after only cases where both
components have a well determined radial velocity, and a relative value in this
quantity of $< 4$km s$^{-1}$, does not show any conspicuous over-densities, not even
along the Galactic plane, which does remain evident in the binary candidates
before the use of the relative radial velocity cut. }

Finally, we take advantage of a CMD selection strategy to further reduce the probability of hidden tertiary systems remaining in the high quality wide binary
samples used. The left panel of Fig. 2 shows a CMD of all the stars in the original 97,251 candidate binaries as grey dots. The stars in the 1,352 sample appear
as black dots, already showing a much more well defined main sequence than the original sample. A small number of photometric binaries remain above this
well-defined main sequence.

We now impose a further cut defined on the CMD, following again the conclusions of Belokurov et al. (2020), who show that unresolved binaries
can be further avoided by remaining within a region of the CMD below the turn-off and excluding the less massive and hence dimmer lower tail of the main
sequence. We hence exclude all binary candidates containing one or two stars lying outside of a region having a G magnitude width of 0.6 magnitudes
about the line connecting points (0.7, 4.7) and (2.2, 9.7) in the (colour, {  absolute} magnitude) plane shown. This last cut leaves only 688 binary
pairs, shown in the right panel of Fig. 2. 

{  The above CMD selection now implies an absolute magnitude limit of 9.7, although only a handful of stars are
actually above an absolute magnitude of 9. Indeed those few stars mentioned above are
amongst the dimmest and with the exception of only one, are excluded by the final signal
to noise cuts in the kinematic plot. Thus, an absolute magnitude cut of 9 is
representative of the final sample. This then becomes a distance-dependant apparent
magnitude cut, which for the mean distance of the final high-quality cut of 90.25 pc
becomes 18.77. For the outer limit of this sample, the corresponding value {  is} 19.5.

When comparing the present final high-quality sample to the final sample in our previous study, only
40$\%$ of the stars in Hernandez et al. (2022) appear in our present sample.
This is due to the significant increase in {\it Gaia} sources having reported radial velocities in going
from eDR3 to DR3, of a factor 4.7. Keeping a sample of comparable size to the one of our
previous study now allows stricter quality cuts. In Hernandez et al. (2022), no binary probability
cut as internally determined by Gaia was applied, as the $B_{p}$ parameter was not available in eDR3.
As discussed above, the present sample includes a $B_{p}<0.2$ cut, with the final sample mean value of this internally
determined {\it Gaia} binarity probability per source, which uses astrometric, photometric and
spectroscopic information in ways complementary to the RUWE filter and the radial velocity
determinations, of $<B_{P}>=0.07$.

To summarise the strategy applied towards the removal of hidden tertiaries, the first cut is through the use of a limit
in the {\it Gaia} DR3 internally determined binary probability parameter of each source,
the $B_{p}<0.2$ constraint described above. The second is through the use of the
RUWE$<1.2$ constraint, this ensures sources where the single star photometric solution
is poor will not be included. Then, keeping only binaries where radial
velocities are reported ensures again a high quality spectroscopic single star
solution resulted for the star in question,  which is easily degraded by
a hidden tertiary. Finally, following Belokurov et al. (2020), a careful
CMD selection strategy is applied. This allows not only the exclusion of clear
photometric binaries, but also limits consideration to regions of the CMD where
errors and photometric blending are minimised. Indeed, through careful statistical
re-sampling of mock observations of simulated hidden tertiaries and then 'observed'
assuming actual Gaia sensitivity parameters, the above authors estimate only a 5\%
hidden tertiary contamination, down to Jupiter scales, when taking a RUWE$<1.4$, plus
the restriction to a well defined region of the CMD, out to 1 kpc, using DR2.
Necessarily, the same CMD criterion as applied here, in conjunction with a RUWE$<1.2$
cut and distances of $<125$pc using the longer base line DR3 catalogue, will 
be highly clear of hidden tertiaries.

Hence it is astrometry, photometry and spectroscopy, {  through} four
different and independent combinations of constraints that are used to eliminate hidden
tertiaries. All cuts are applied sequentially to both components of each candidate
binary, with all such candidates where even one of the components fails the cut being
removed.
}

Before presenting the relative velocity vs. internal separation and relative velocity vs. mass scalings of our final samples, the following section details
the mass estimates used.

\section{Addressing Mass estimate biases}

\begin{figure*}
    \includegraphics[height=7.0cm,width=8.8cm]{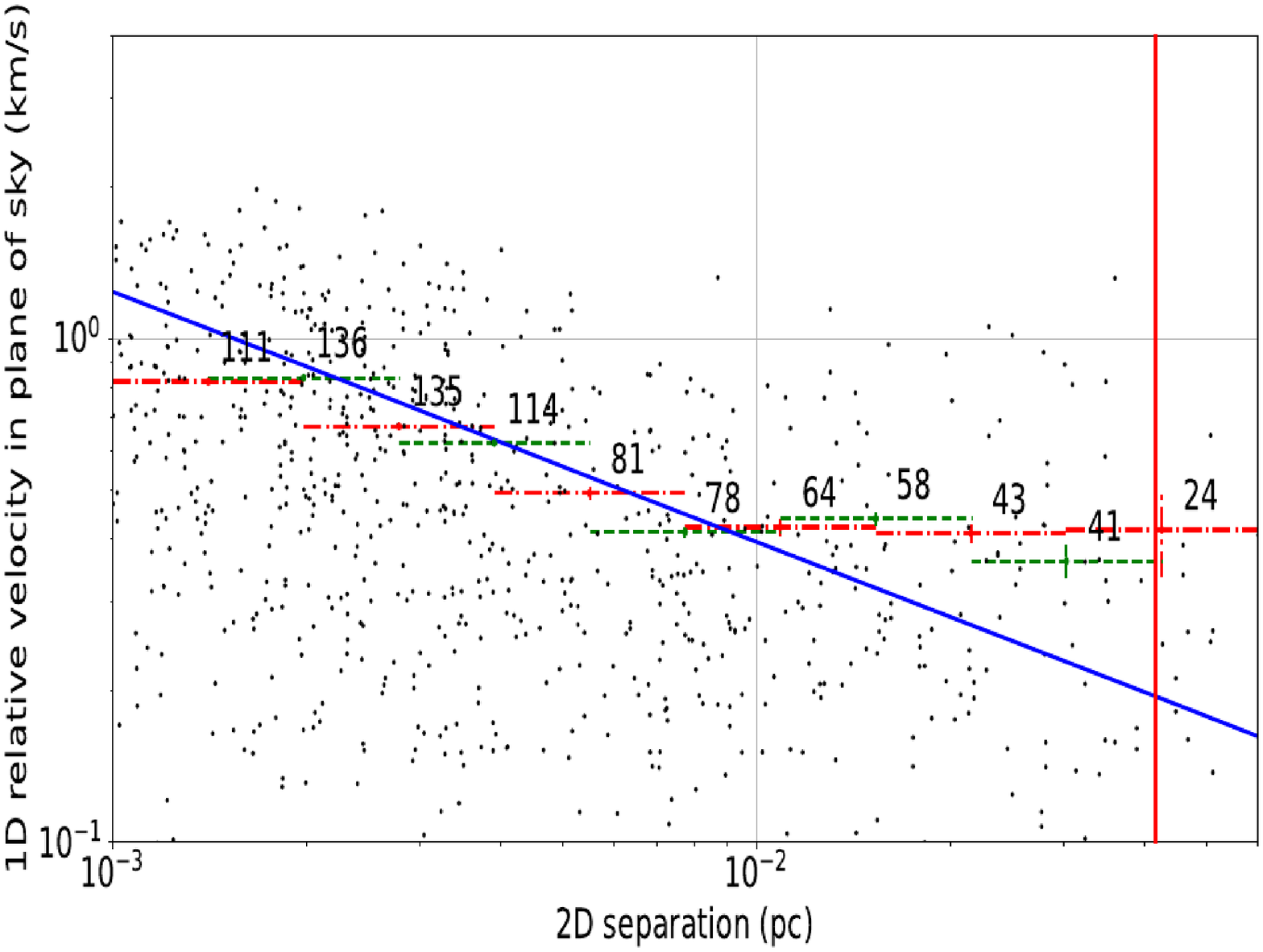}
     \hspace*{-5pt}
     \includegraphics[height=7.0cm,width=8.8cm]{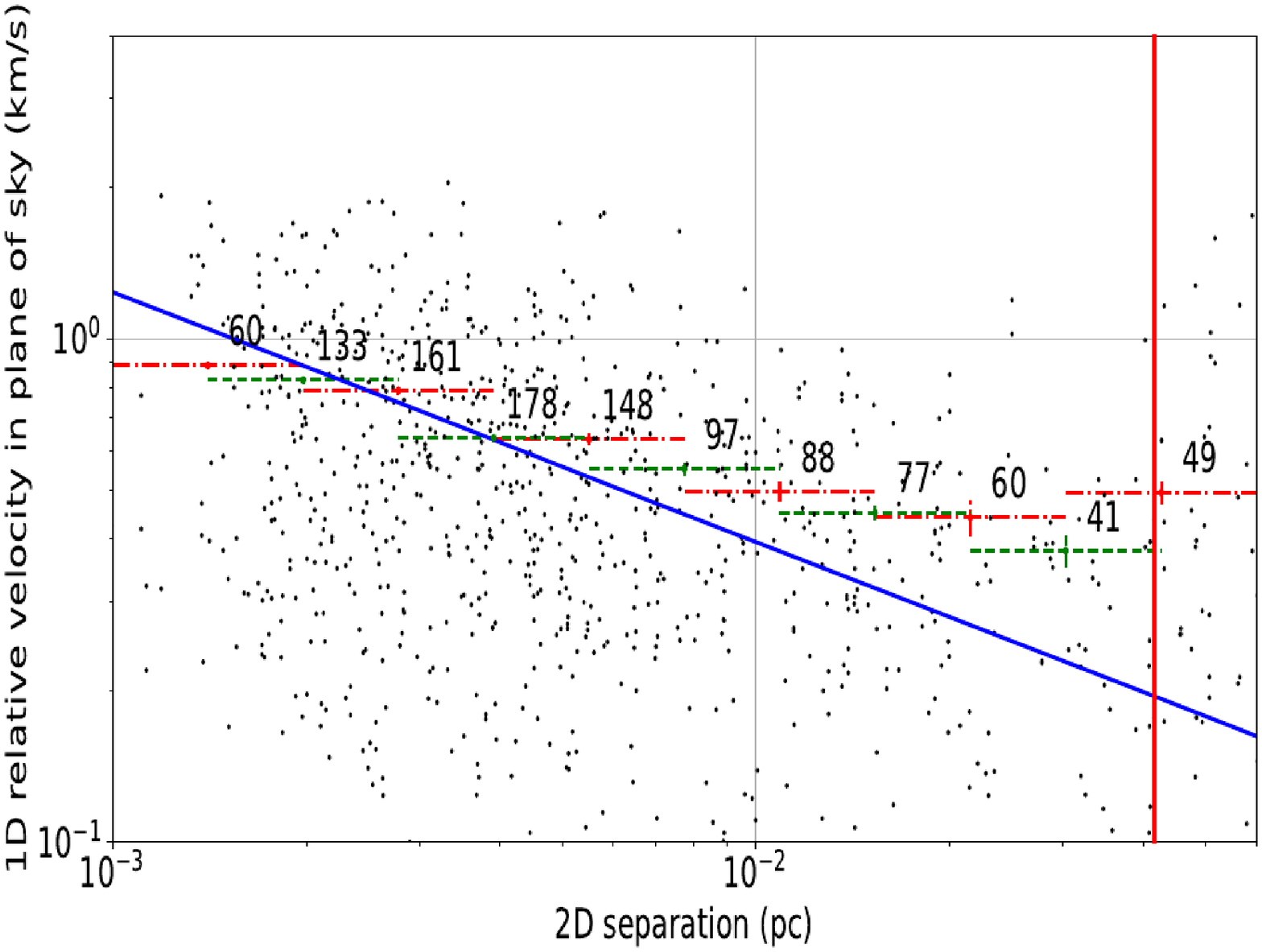}
    
     \caption{{  Left(a)}:Kinematic plot showing 1D $\Delta V$ values on the plane of the sky, dots, for binaries in the $(D/$pc$)<125$ pc sample as a
       function of the internal separation of each binary in pc, $s$, R.A. and Dec values appearing separately for each binary. The points with error bars give
       the binned rms values for this same quantity, with the horizontal bars giving the bin size and the vertical ones 1$\sigma$ confidence intervals on
       the quantity being plotted, dashed and dot-dashed bins for R.A. and Dec. observations, respectively. The diagonal line gives the prediction for the
       binned rms 1D $\Delta V$ values on the plane of the sky for solar neighbourhood wide binaries from Jiang \& Tremaine (2010), and the vertical one
       shows the MOND radius for the mean total mass of the binaries shown. {  Right(b)}: Same as
       the left panel, but for the $125<(D/$pc$)<170$ sample, see text and Table 1 for all other signal-to-noise, RUWE, $B_{p}$ and CMD diagram selection cuts,
       which were identically applied to both distance samples shown in this figure.}   
 \end{figure*}

Determining if the internal kinematics of a sample of wide binaries is {  consistent} with Newtonian gravity, clearly requires estimating the masses of each
star in each of the binaries being studied {  (e.g. Banik \& Zhao 2018, Pittordis \& Sutherland 2023)}. In the context of large {\it Gaia} samples of Solar
Neighbourhood wide binaries, this mass estimate has so far been approached in terms of simple magnitude-mass scalings, such as:

\begin{equation}
\left( \frac{M}{M_{\odot}} \right) = 10^{0.0725(4.76-M_{G})},
\end{equation}

\noindent expected to be reasonably accurate and unbiased for the old low mass stars of the Solar Neighbourhood comprising the relevant wide binaries e.g.
Pittordis \& Sutherland (2019), Hernandez et al. (2022). However, the availability of an internally determined {\it Gaia} FLAME work package mass estimate
making use of spectroscopic information in DR3, allows for a much more accurate update in the masses of the binaries analysed. Unfortunately, the more
demanding observations required for {\it Gaia} spectroscopic mass determinations imply that this parameter is available only for a fraction of even nearby stars.
If we restrict ourselves to wide binaries where both components have a {\it Gaia} DR3 mass determination, we would have to eliminate all candidates where no
such information is available, and also those cases where {\it Gaia} DR3 masses are available for only one of the two members of a binary. Therefore, 
supplementing internally determined spectroscopic {\it Gaia} DR3 masses with estimates from eq.(1), in cases where the former is not available, is desirable.

The first test of this scheme is to check for the presence of any biases or inconsistencies when comparing the results of eq.(1) to {\it Gaia} DR3 masses. I begin
with a sample as described in the previous section, taking a distance cut of $D<135$ {  pc}, and changing only the $RUWE$ and $B_{P}$ cuts to $<2.0$ and $<0.4$.
This relaxation of the quality controls only at this point, serves to increase the sample so as to obtain a more complete comparison of the two mass estimates
mentioned above. This sample yields 3,696 binary systems.

The left panel in Fig. 3 shows the individual stars of the sample just described which have {\it Gaia} DR3 masses, in a plot comparing this quantity to the result
of eq.(1) for these same stars, in the mass range relevant for the wide binaries of interest. The grey dots show stars which are members of a binary system where the
other star does not have a {\it Gaia} DR3 mass determination, and the black dots cases where the other component of the binary system also has a {\it Gaia} DR3 mass
available. As can be seen, a significant fraction of binaries do not have {\it Gaia} DR3 masses for both components, indeed, in the sample shown only $24\%$ of the
stars have {\it Gaia} DR3 masses available for both components. Thus, given the small available high quality samples, it is important to supplement {\it Gaia} DR3
mass determinations with a reliable estimate inferring masses from {\it Gaia} DR3 quantities available for a larger fraction of stars. Unfortunately, we see from
the left panel in Fig. 3 that the masses inferred using equation (1) are not an unbiased estimate of the {\it Gaia} DR3 masses. For masses below about
$0.7 M_{\odot}$ a small systematic appears such that the inferences from eq.(1) are on average $10\%$ smaller than the more accurate {\it Gaia} DR3 ones which
use more detailed spectroscopic information.

For masses $>0.7 M_{\odot}$, mass estimates from eq.(1) and {\it Gaia} DR3 agree much better and show almost no systematic biases. However, stars with masses in the
range where both inferences diverge form a large fraction of the constituent members of the relevant binaries, e.g. Hernandez et al. (2022) report a mean binary
mass of $m_{B}=1.6 M_{\odot}$ for their {\it Gaia} eDR3 sample, and in the study by Pittordis \& Sutherland (2019), inferred  mass distributions for individual stars
peak for $0.4 M_{\odot}<m_{\star}<0.5 M_{\odot}$. Indeed, attempting to test particular MOND models requires masses determined to close {  to $10 \%$ accuracy }
(Banik \& Zhao 2018), and crucially, no systematic biases in this quantity, making the need of a non-biased mass estimate for all stars involved an important
requirement in the context of wide binary gravity tests.

To this end we perform a simple correction on eq.(1), adding $0.05 M_{\odot}$ to the mass of all stars which eq.(1) returns as being below $0.7 M_{\odot}$. The result
of this adjustment is shown in the right panel of Figure 3, where it is clear that both mass estimates are now in much better agreement and no substantial systematic
bias remains. A more detailed adjustment is of course possible, but will only yield much smaller corrections, which in any case will eventually be rendered unnecessary
as the fraction of accurate {\it Gaia} spectroscopically determined masses grows with time. For the remainder of this paper, I shall use {\it Gaia} DR3 masses
when available, supplemented by the use of eq.(1), with the inclusion of the correction described, in cases where no {\it Gaia} DR3 masses exist for any particular
star. This is in effect equivalent to the more detailed higher order mass-magnitude scalings used by other authors, e.g. Chae (2023).

\begin{figure*}
    \includegraphics[height=7.0cm,width=8.8cm]{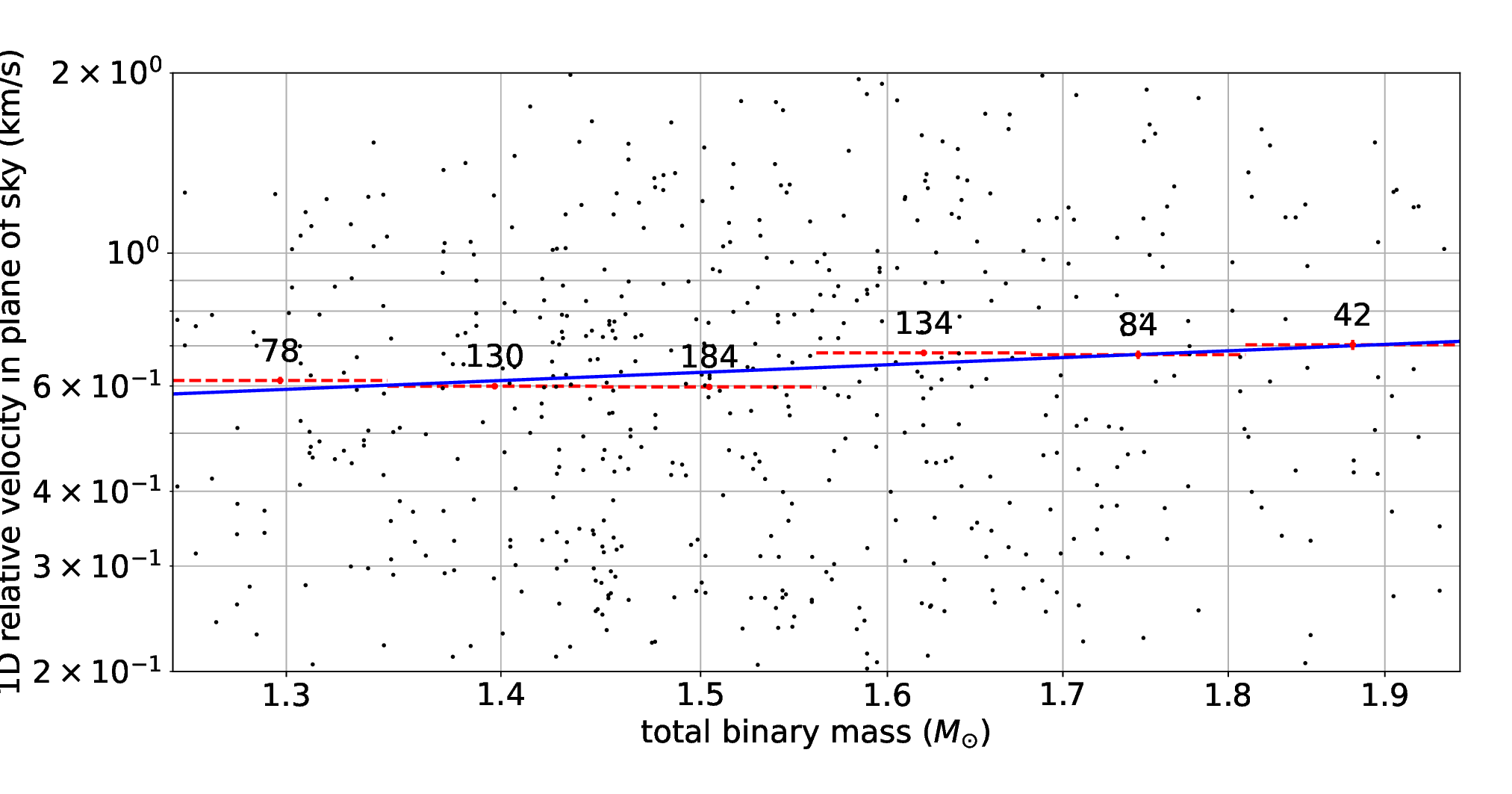}
     \hspace*{-5pt}
     \includegraphics[height=7.0cm,width=8.8cm]{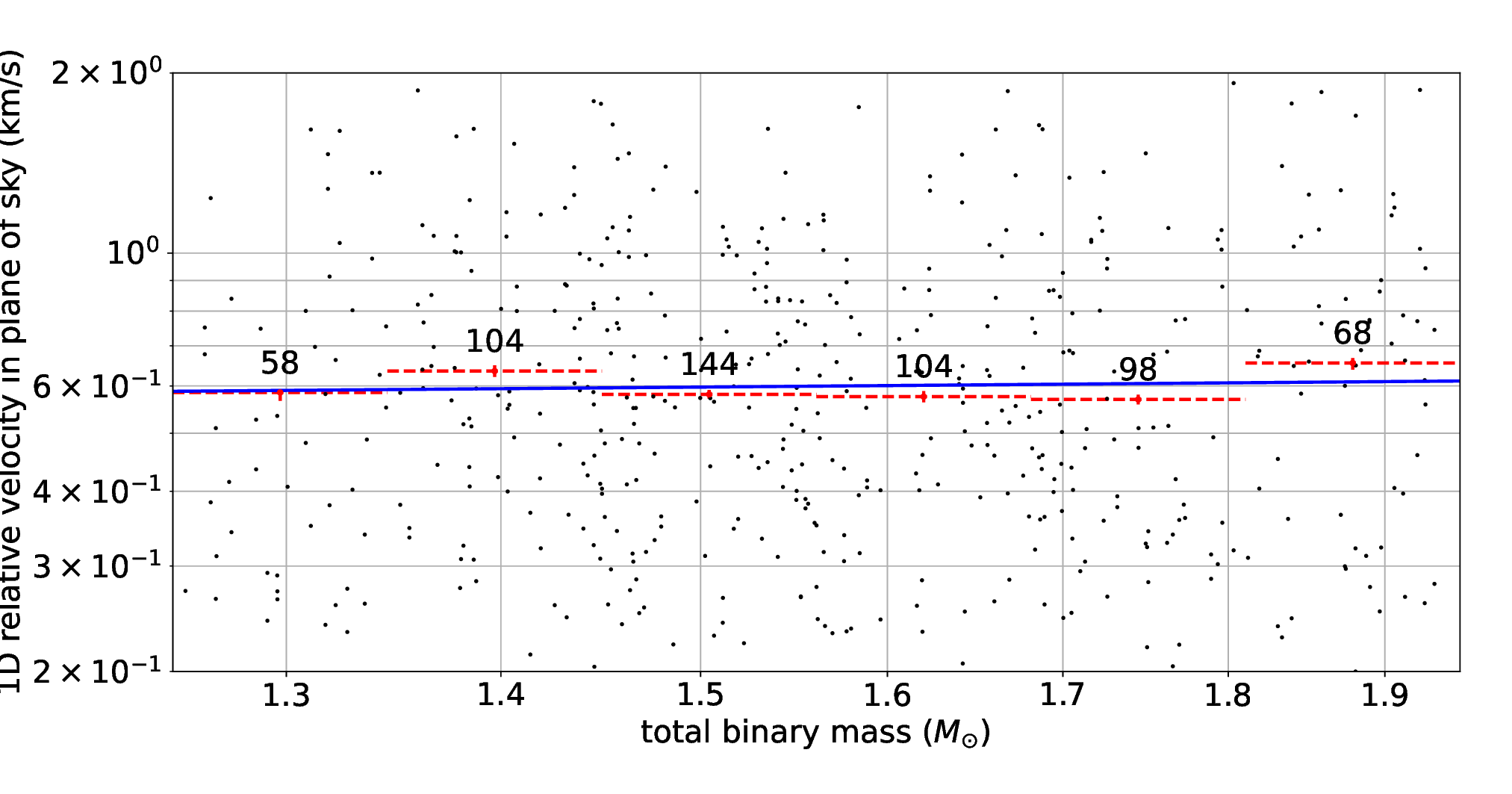}
    
     \caption{{  Left(a)}:log-log plot of 1D $\Delta V$ values on the plane of the sky, dots, for binaries in the $(s/$pc$)<0.01$, $(D/$pc$)<125$ sample as
       a function of total binary mass, R.A. and Dec values appearing separately for each binary. Dots with error bars give the mean values of said quantity per bin,
       together with 1$\sigma$ confidence intervals on these means. The solid line shows a linear fit to the binned means having a slope of $\alpha=0.46 \pm0.13$
       and giving a correlation parameter of $r=0.87$, consistent with Newtonian expectations in this high acceleration region. {  Right(b)}:Same as the left panel,
       but for the $(s/$pc$)<0.01$, $125<(D/$pc$)<170$ sample. The solid line shows a linear fit to the binned means having a slope of $\alpha=0.089 \pm0.21$ and
       giving a correlation parameter of $r=0.21$, {  barely consistent with Newtonian expectations in this high acceleration region at a $2\sigma$ level},
       and indicating some combination of error dominated $\Delta V$ determinations and the presence of kinematic contaminants in this more distant sample.}   
 \end{figure*}

\section{Results}

Starting from the distance cut described at the end of section 2 with $D<125${  pc} leaves 688 binary pairs, after the CMD cut criteria presented there. For each
of these binaries, parallaxes, positions and proper motion {\it Gaia} DR3 parameters are used to evaluate the relative velocities on the
plane of the sky in R.A. and Dec., $(\Delta V)_{RA}$ and $(\Delta V)_{Dec}$, including spherical geometric effects. {  We consider only velocity
differences on the plane of the sky for the kinematic tests performed because these are more robust against contamination from the presence of hidden
tertiaries than velocities along the line of sight. Velocities on the plane of the sky are being inferred through proper
motions, while along the line of sight, velocities are derived through the Doppler effect. This is important since a hidden tertiary of short period could
result in an added velocity component, which if along the line of sight, will directly result in an equal kinematic contaminant on the radial velocity,
regardless of the amplitude of the stellar oscillation. However, on the plane of the sky, the same velocity contaminant, for a very tight hidden tertiary
having a small internal semi-major axis, will not result in any kinematic contamination if the resulting oscillation is of small spatial amplitude.
Distortions of small spatial amplitude (even those of high velocity) will to a large degree be averaged out over the 34 month timeline of the DR3 catalogue and
produce negligible effects on plane of the sky velocities, whenever the inner orbital periods are significantly shorter than the integration period of the
catalogue.}

Also, restricting the relative velocity analysis to the plane of the sky (as was also done and clearly stated in Hernandez et al. 2022), makes this study
robust to general relativistic gravitational redshift effects distorting relative velocities along the line of sight {  (e.g. El-Badry 2022)},
the opposite of what was mistakenly claimed by Loeb (2022). Our kinematic $\Delta V$ determinations include only relative motions on the plane of the sky
{  determined primarily through {\it Gaia} proper motions, positions and parallaxes; radial velocities play only a very minor part, exclusively in the
small perspective corrections, which as shown by El-Badry (2019), are second order for separations below 0.1 pc.}

Once the $(\Delta V)_{RA}$ and $(\Delta V)_{Dec}$ parameters are obtained for the binaries in question, two final quality cuts are introduced. The first is
the exclusion of low signal-to-noise cases, I keep only binaries where both $\Delta V$ measurements satisfy $(\Delta V/\sigma_{\Delta V})>1.5$. Lastly,
I evaluate the final average binary probability {\it Gaia} parameter for the sample, $<B_{P}>$, and remove half of that number as a percentage of the
highest $\Delta V$ systems, uniformly distributed along the separation, $s$, interval considered. The logic for this last cut is that if say, $<B_{P}>=0.07$,
it is possible that $7 \%$ of our remaining systems harbour a hidden tertiary. Some of those will have negligible effects on the resulting $\Delta V$
measurements of the binaries, when the hidden tertiary velocity perturbation lies primarily along the line of sight, but in some cases this kinematic
contamination will occur substantially on the plane of the sky and hence distort the intended test. There is no guarantee that these last cases will be
those showing the largest $\Delta V$ values across our sample, but as a measure to ensure the least possible effect of any degree of contamination from
hidden tertiaries, the fastest $3\%$ of cases per separation bin are {  eliminated} across the entire $s$ range covered.

\begin{figure*}
    \includegraphics[height=7.0cm,width=8.8cm]{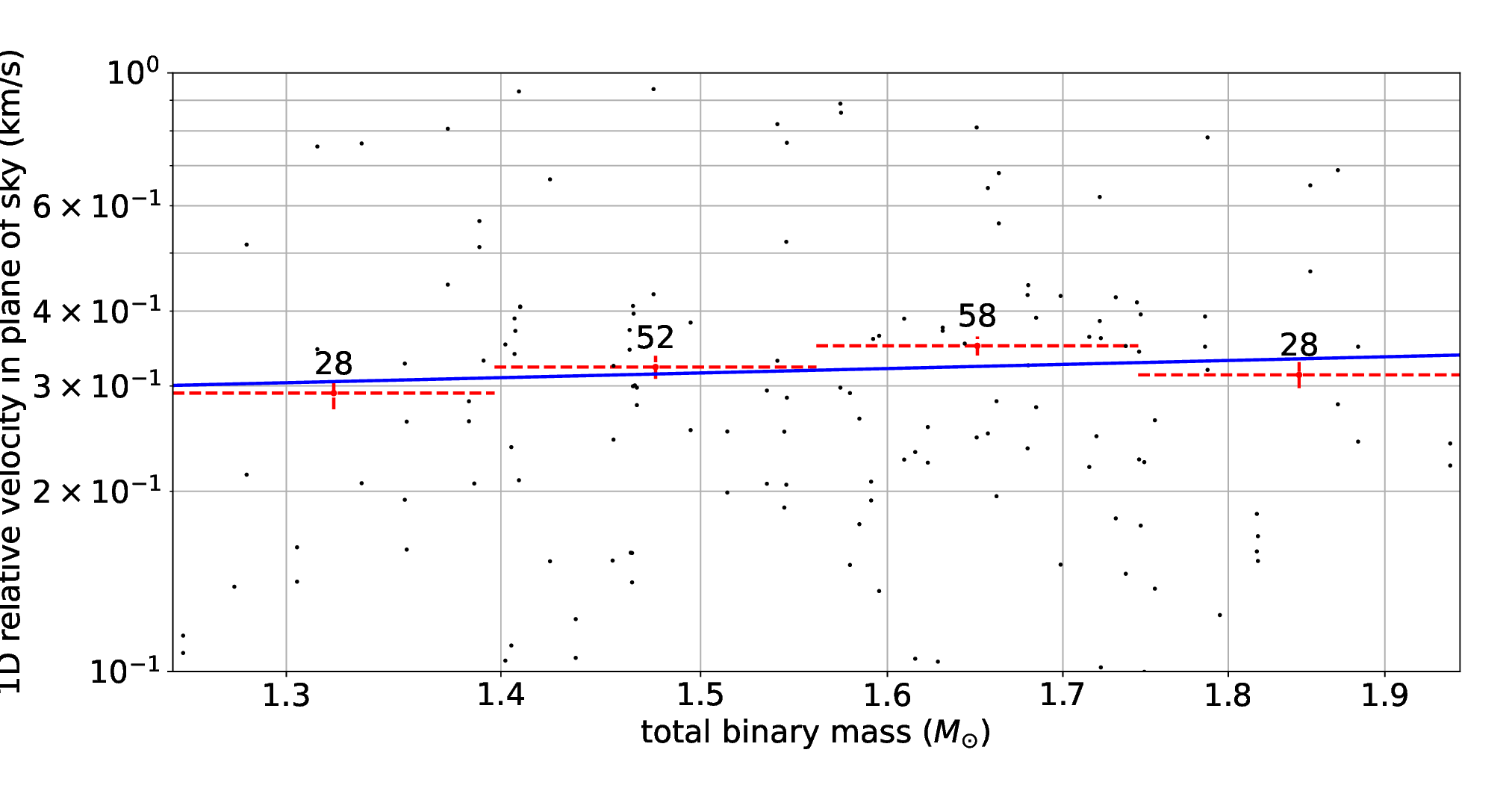}
     \hspace*{-5pt}
     \includegraphics[height=7.0cm,width=8.8cm]{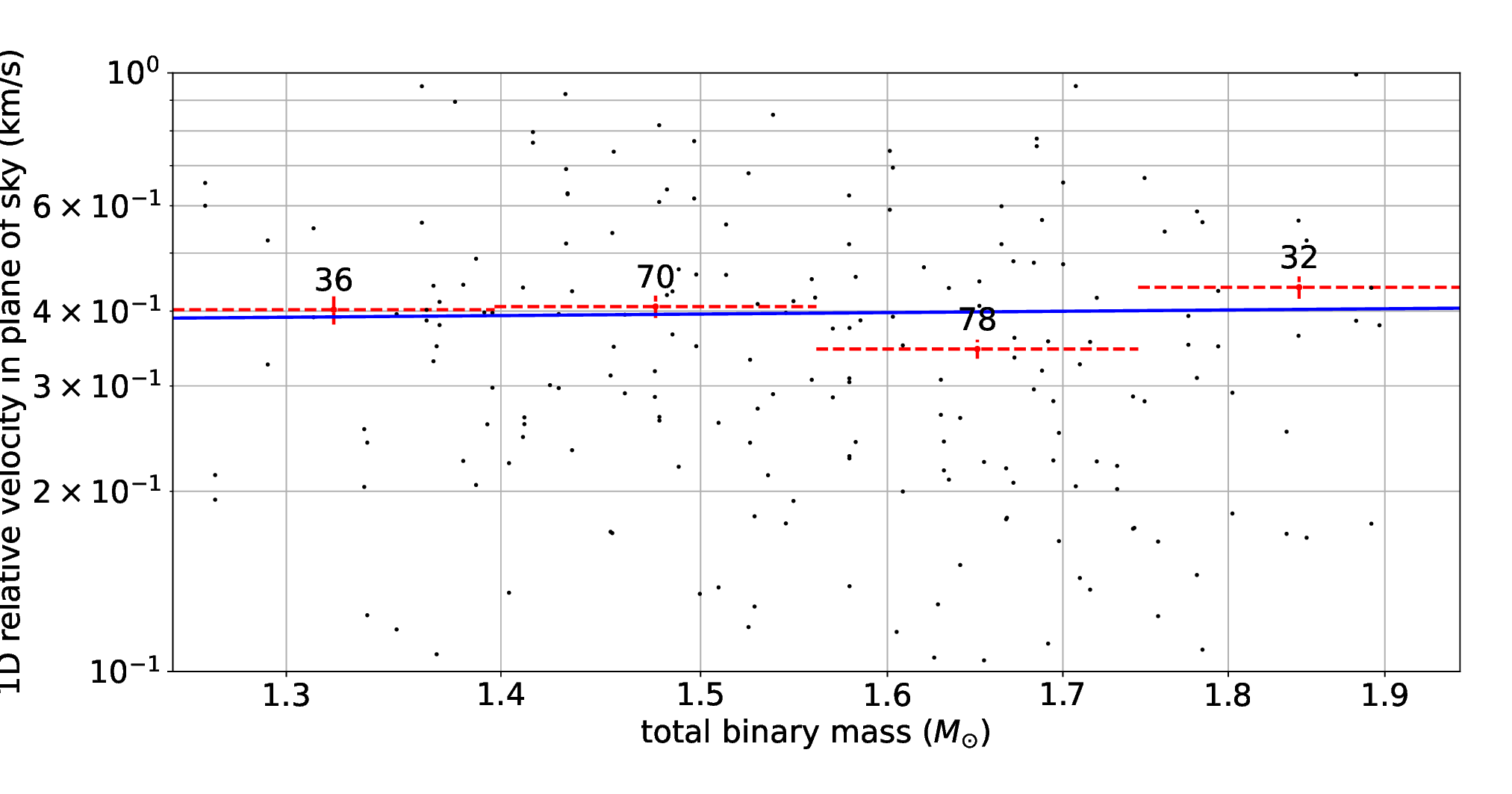}
    
     \caption{{  Left(a)}:log-log plot of 1D $\Delta V$ values on the plane of the sky, dots, for binaries in the $(s/$pc$)>0.01$, $(D/$pc$)<125$ sample as a
       function of total binary mass, R.A. and Dec values appearing separately for each binary. Dots with error bars give the mean values of said quantity per bin,
       together with 1$\sigma$ confidence intervals on these means. The solid line shows a linear fit to the binned means having a slope of $\alpha=0.263 \pm0.32$
       and giving a correlation parameter of $r=0.50$, consistent in this low acceleration region with {  the $1/4$ of {  Tully}-Fisher kinematics, and also
       with the $1/2$ of Newtonian expectations, at a $1\sigma$ level}. {  Right(b)}:Same as the left panel,
       but for the $(s/$pc$)>0.01$, $125<(D/$pc$)<170$ sample. The solid line shows a linear fit to the binned means having a slope of $\alpha=0.086 \pm0.49$ and
       giving a correlation parameter of $r=0.12$, {  consistent with either Newtonian or Tully-Fisher expectations in this low acceleration region, but
       at a low significance level, given the large confidence intervals, probably} indicating some combination of error dominated $\Delta V$ determinations
       and the presence of kinematic contaminants in this more distant sample.}   
 \end{figure*}

The two final quality cuts mentioned above leave a small and very pure sample of 450 highly isolated wide binaries, a scatter plot of the $(\Delta V)_{RA}$ and
$(\Delta V)_{Dec}$ values obtained is presented in the left panel of Fig.4, as a function of the internal separation on the plane of the sky for each binary.
{  In essence, each binary has been observed along two perpendicular directions, yielding two 1D $\Delta V$ observations which are independent from the point
of view of projection effects. The robustness of the results obtained to this assumption, also included in all previous published wide binary papers by this
group, is tested and confirmed in the appendix.} The final average distance to these stars {  is} $90.25$ pc, with the selection criteria used resulting in
very high average signal-to-noise for relative $\Delta V$, stellar proper motion and parallax of 15.71, 3442 and 855.4 respectively, and average $RUWE$ and $B_{P}$
values of 1.01 and 0.07 respectively, as detailed in Table 1.

The points with error bars give the binned rms values and corresponding $1\sigma$ confidence intervals for R.A. and Dec. measurements, dashed and dot-dashed cases
respectively, where the horizontal lines are just the bin sizes and the figures above the bins give the number of data points per bin. This can be compared to
detailed Newtonian expectations for 1D relative velocities between components of wide binaries in the Solar Neighbourhood through the work of Jiang \& Tremaine
(2010), henceforth JT10. These authors simulate populations of 50,000 wide binaries in the Solar Neighbourhood, evolved over a 10 Gyr period under the influence
of Galactic tides and encounters with field stars and molecular clouds, under the assumptions of Newtonian dynamics, constant total binary masses of
$m_{B}=2 M_{\odot}$ and expected distributions of ellipticities and isotropic orientations with respect to a simulated observational line of sight. The final rms
value of the 1D $\Delta V$ measurements was then reported, and is shown by the solid line in the left panel of Fig. 4. In the $s$ range given we see essentially
a Keplerian $\Delta V \propto s^{-1/2}$ scaling, as we remain {  well within the $\approx 1.7$ pc Jacobi radius} of the problem.

{ 
As explained above, an important aspect of the wide binary test is to consider 
{  projection} effects and a distribution of ellipticities, even
within the Newtonian region. I have not performed any dynamical modelling, neither
within a Newtonian model nor within any modified gravity one. The main intent is only to
test the validity of a full Newtonian model. To this end, the consequences of projection
effects and of a distribution of ellipticities under a Newtonian scheme were taken into
account through a detailed comparison to the results of JT10. In
that paper the authors simulate large populations of wide binaries, and provide detailed
predictions for the 1D projected rms values of the resulting relative velocities as a
function of projected separations. It is for this reason that those are precisely the
quantities calculated here, as it is only those quantities which can be compared in detail
to the predictions of JT10. A further important point to note is that
JT10 do not present initial instantaneous predictions
for the rms value of binned 1D relative velocities as a function of projected
separations, but those quantities after 10 Gyr of dynamical evolution in the tidal
field of the Solar Neighbourhood and under the dynamical influence of gravitational
interaction {  of} the wide binary populations treated with field {  stars}. This is important,
as the results of this evolution alter the final distributions of ellipticities away
from the initial assumed ones. One should not compare against birth distributions of
ellipticities, $e$, but against the expected present day ones. To this effect, note that
JT10 assume an initial uniform distribution of $e^{2}$ values between 0 and 1, which is known as
a thermal distribution of ellipticities and is justified in terms of thermodynamical expectations
in the sampling of the energy distribution of initial wide binary parameters, e.g. Kroupa (2008).
However, recent direct inferences of $e$ distributions by  Hwang et al. (2022) using {\it Gaia}
data of local wide binaries, show that the present-day distribution of ellipticities becomes super-thermal
in going to the $s>5\times 10^{-3}$pc region, precisely the region of relevance to the wide binary test.
}

We see that in the separation range $s<0.01$ pc {  (2000 AU)}, results closely follow the predictions of JT10, confirming
their Newtonian expectations for the rms values of 1D $\Delta V$. Our inferred values lie slightly below the Newtonian predictions,
in consistence with the final sample $<m_{B}>=1.56 M_{\odot}$ being a factor of $0.78$ below the assumed $m_{B}=2 M_{\odot}$ of JT10.
{  Indeed, our results are below their prediction} by a factor of very close to $0.78^{1/2}=0.88$.

\begin{table*}
\begin{flushleft}
  \caption{Parameters for the two distance selection cuts described.}
  %  \begin{tabular}{@{} | l | llll  | @{}}
  \begin{tabular}{ | l | c c c c c c c c c | }
  \hline
  \hline   
 Distance    & Initial No.  & No. after & No. clearing           & $<D/pc>$   & $<S/N>_{\Delta V}$ & $<S/N>_{pm}$ &  $<S/N>_{\varpi}$  & $<RUWE>$ & $<B_{P}>$      \\
 selection   & of binaries  & CMD cut  & $(\Delta V/\sigma_{\Delta V})>1.5$,  &  &              &             &                  &          &                \\
             &              &          & $<$top $3\% \Delta V$              &  &              &             &                  &          &                \\   
   \hline
 $(D/pc)<125 $    & $1352$       &  $688$    &  $450$                 & $90.25$ & $15.7$          & $3442$       & $855.4$          & $1.01$    & $0.07$        \\
 $125<(D/pc)<170$ & $1562$       &  $914$    &  $450$                 & $147.6$ & $7.5$           & $1798$       & $474.9$          & $1.01$    & $0.06$        \\

 \hline 
 
\end{tabular} 

  For the two distance cuts described in the text, $(D/$pc$)<125$ and $125<(D/$pc$)<170$, the first three entries of the table give the initial number of
  $(S/N)_{\varpi} >100$, $RUWE<1.2$, $B_{P}<0.2$ binaries after de-grouping, the remaining binaries after application of the CMD filtering procedure,
  and the final number of binaries in the kinematic plots after removal of $(\Delta V/\sigma_{\Delta V})<1.5$ cases, and the exclusion of the $3\%$ fastest binaries
  per bin. The following two entries show the final average distance in pc and $\Delta V$ signal-to-noise ratio, averaged over R.A. and Dec, for the binaries
  in the final samples. The last four entries give the average signal-to-noise values in proper motions (averaged over R.A. and Dec.) and parallax, and final
  mean $RUWE$ and $B_{p}$ values for all the individual stars included in the kinematic plots of figure 4.

\end{flushleft}
\end{table*}

The situation is qualitatively different in the $s>0.01$ pc region, where the rms {\it Gaia} DR3 inferred $\Delta V$ values cease to follow the Newtonian
$\Delta V \propto s^{-1/2}$ scalings and settle to a constant value with distance. It is interesting that this asymptotic value of slightly below $0.4$ km s$^{-1}$
closely agrees with the spiral galaxy baryonic Tully-Fisher velocity scaling {  of $V_{TF}=(GMa_{0})^{1/4}=0.35(M/M_{\odot})^{1/4}$ km s$^{-1}$, which} when evaluated
at a total baryonic mass of $1.56 M_{\odot}$ yields $0.39$ km s$^{-1}$.

{  An indicative scale at which MOND phenomena are expected for a system of total mass $M$, is given by the MOND radius, $R_{M}=(GM/a_{0})^{1/2}$,
the radius at which the acceleration of a test particle in orbit about a spherically symmetric mass distribution becomes equal to $a_{0}$. For the
average total binary mass of the sample in Fig. (4), $<m_{B}>=1.56 M_{\odot}$ and $<m_{B}>=1.59 M_{\odot}$, left and right panels respectively,
$R_{M}=0.042$ pc (8400 AU). This value is shown by the red vertical line in Fig. (4)}. Comparing to the $0.01$ pc (2000 AU) separation where we see the
regime transition, would leave a factor of 4.2 to account between a form factor due to the difference in symmetry conditions between circular equilibrium
orbits and the highly dynamical problem of populations of two orbiting bodies with a distribution of effective eccentricities, and the appearance of
gravitational anomalies {  at accelerations} a little above the $a<a_{0}$ threshold. Indeed, such is the case in spiral galaxy dynamics, e.g. at the
Solar Radius $a\approx a_{0}$ and the internal dark matter fraction required under Newtonian gravity is already about 0.5. At the Solar circle radius of
about 8.2 kpc, the amplitude of the Newtonian baryonic rotation curve of the Galaxy is about 185 km s$^{-1}$ but the actually observed rotation velocity
is about 230 km s$^{-1}$. Thus, gravity is about 1.5$\times$ stronger than the Newtonian expectation with only baryons, implying a dark matter to baryon
ratio of about 0.5, see e.g. Zhu et al. (2022).

A finer test geared towards understanding the physics behind the $\Delta V$ vs. $s$ scalings discussed above comes from plotting the $\Delta V$ observations
against the total binary mass of each system, $m_{B}$. For the sample being discussed, this appears in the left panel of Fig. 5, including only binaries
within the internal separation interval $s<0.01$ pc, the region consistent with a Newtonian $\Delta V \propto s^{-1/2}$ scaling, shown in the left panel of Fig. 4.
The dots with error bars now give the average $\Delta V$ values for binned binaries, with R.A. and Dec. treated as independent data points. The horizontal
lines give the width of the bins, while the vertical ones indicate the $1\sigma$ confidence intervals on  $<\Delta V>$ and the numbers above show the number
of data points per bin. A linear fit to the binned $<\Delta V>$ values gives a scaling of $<\Delta V>\propto m_{B}^{0.46 \pm 0.13}$ with a correlation parameter of
$r=0.87$, perfectly consistent with Newtonian expectations of $<\Delta V> \propto m_{B}^{0.5} $.

The corresponding plot for the $s>0.01$ pc region of the left panel in Fig. 4 is given in the left panel of Fig.6. This time the linear fit gives
$<\Delta V>\propto m_{B}^{0.26 \pm 0.32}$. This scaling, though still consistent with Newtonian expectations at a $1\sigma$ level, is in closer agreement
with a galactic baryonic Tully-Fisher scaling of $<\Delta V>\propto m_{B}^{0.25}$. The narrow mass interval available and the small number of points limit
the precision of this last test, in spite of which a strong reduction in the best fit  $<\Delta V> \propto m_{B}^{\alpha}$ value of $\alpha$ is apparent
on crossing the $s=0.01$ {  pc (2000 AU)} binary separation threshold which divides the region consistent with Newtonian expectations in the previous
$\Delta V$ vs. $s$ plot, from the flat rms relative velocity regime appearing for $s>0.01$ pc in that plot. Notice that not all points considered in the
fits in Figs. 5, 6 appear in the plots, given the $\Delta V$ range displayed.

All parameters of the fits discussed appear in Table 2, including the {  percentage} of binaries used where both masses were supplied directly by {\it Gaia} DR3,
and the percentage of binaries where at least one mass was provided in such a way. For the $s<0.01$ pc region, the above two values are $40\%$ and $61\%$
respectively, with the corresponding values for the $s>0.01$pc region being $41\%$ and $63\%$. We see no significant difference between these two sets of numbers,
showing no bias in the availability of {\it Gaia} DR3 {  mass} determinations, and hence no bias in the availability of high quality data, across the two regions
being compared.

\begin{figure*}
    \includegraphics[height=7.0cm,width=8.8cm]{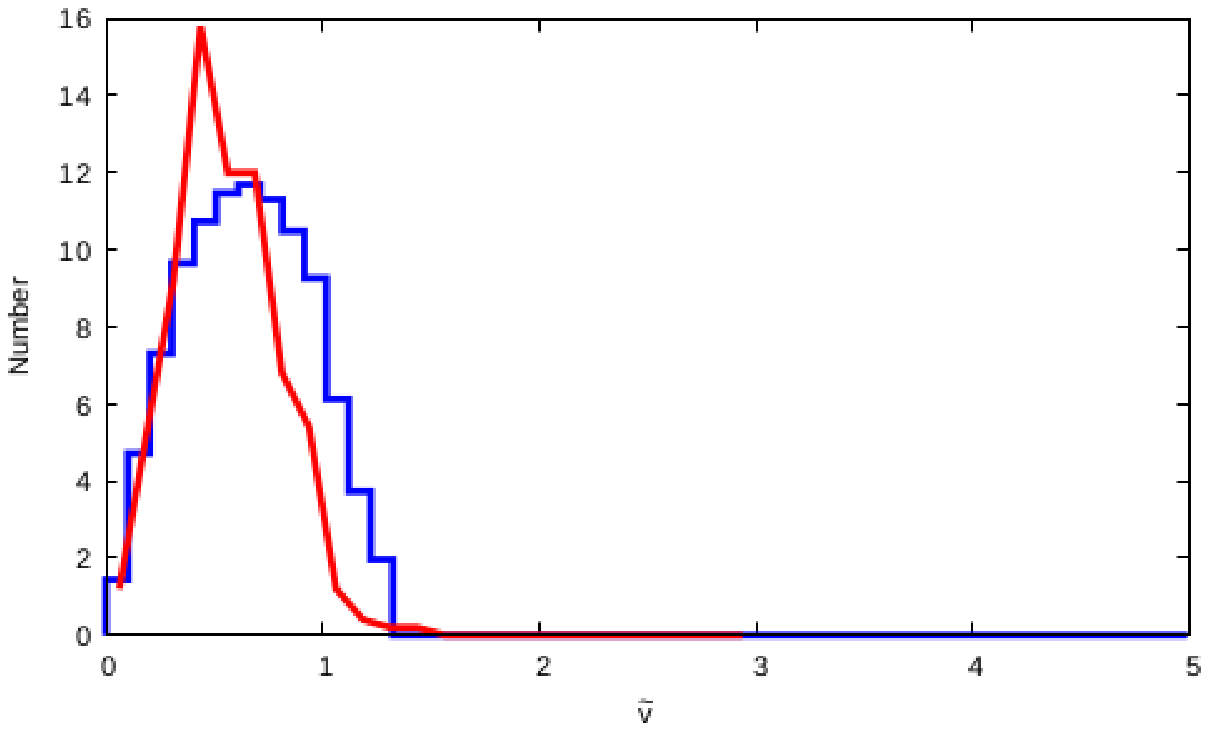}
     \hspace*{-5pt}
     \includegraphics[height=7.0cm,width=8.8cm]{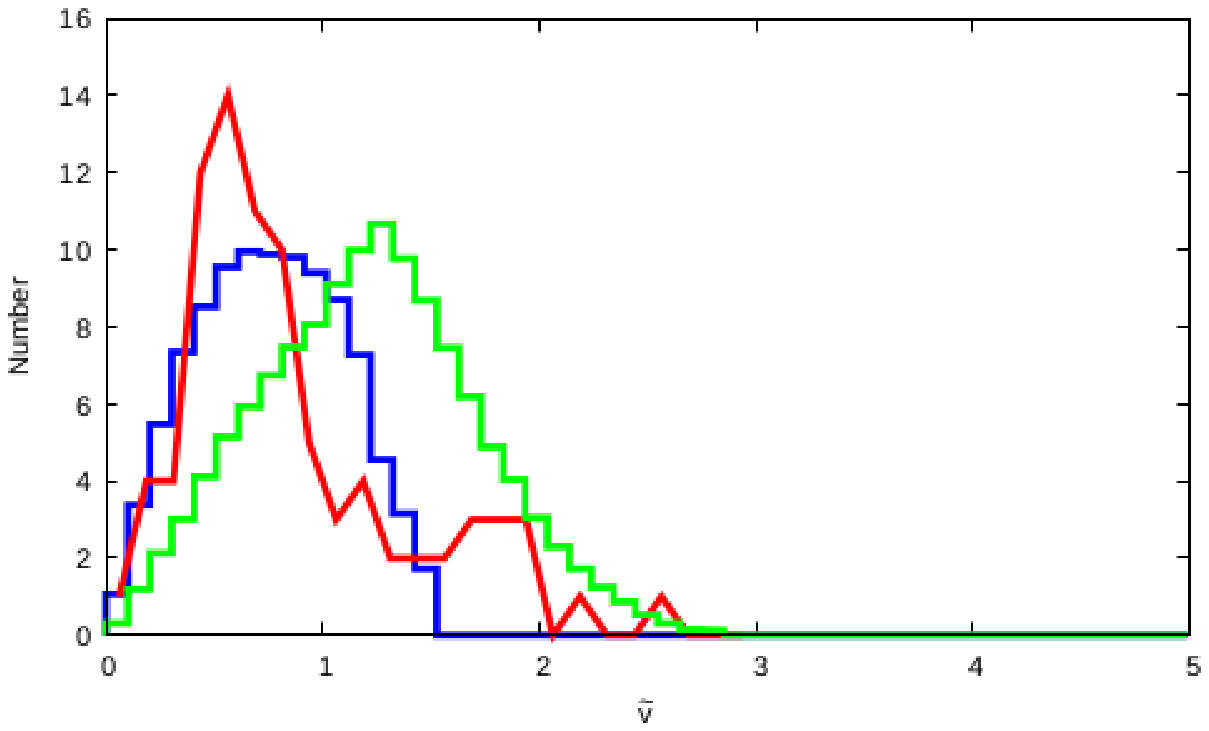}
    
     \caption{{  Left(a)}:Distribution of $\tilde{v}$ values for the wide binaries in the left panel of Fig. 4, for the $s<0.01$pc Newtonian region, red
       histogram. The blue histogram gives Newtonian predictions for this distribution from Banik \& Zhao (2022). 
       {  Right(b)}:Distribution of $\tilde{v}$ values for the wide binaries in the left panel of Fig. 4, for the $s>0.01$pc low acceleration region,
       red histogram. The blue histogram gives predictions for the distribution of $\tilde{v}$ values under an AQUAL MOND model and the green one corresponding
       predictions for a MOND without external field effect model, both from Banik \& Zhao (2022).}   
 \end{figure*}

Despite the extensive pruning of the sample and all the various independent checks introduced to limit the presence of any kinematic contaminants,
it is of course possible that some remain. To gauge the possibility of this, we explore the distance dependence of our results, given that all
errors and kinematic contaminants necessarily grow in importance as the mean distance of the binaries considered increases. With increasing distance
the fixed $(S/N)_{\varpi}>100$ constraint used implies a larger confidence interval along the line of sight and hence a greater probability of including
projection interlopers, while on average stars will appear dimmer, leading to an unavoidable increase in the error intervals for the inferred proper
motions and hence a noisier $\Delta V$ sample. Further, any hidden tertiary induced wobble will more easily hide below more uncertain stellar parameters.

I test for the sensitivity of the results to distance by repeating the experiment described previously, leaving all quality and kinematic contaminant exclusion
cuts the same, but increasing this time the distance cut-off of the sample, to produce a second binary sample independent of the previous one. This time the
distance cut is $125<(D/pc)<170$. The $\Delta V$ vs. $s$ plot of this new sample is shown in the right panel of Fig. 4, while the $s<0.01$ pc $\Delta V$
vs. $m_{B}$ plot appears in the right panel of Fig. 5, the corresponding  $s>0.01$ pc $\Delta V$ vs. $m_{B}$ scaling {  is} shown in the right panel of Fig. 6.
From the sample parameters given in Table 1 we see that the mean distance of the sample has increased by $64\%$, resulting in important decreases in the final
average signal-to-noise values for binary $\Delta V$, stellar proper motion and  parallax of $52.3\%$, $47.8\%$ and $44.5\%$, respectively.

Such important decreases in the quality of the sample inevitably lead to noisier $\Delta V$ data where all trends will be less clear, independent of the
probable increased appearance of kinematic contaminants. The right panel of Fig. 4 no longer shows such a close correspondence to the Newtonian expectations
as seen in the left panel, rms values for $\Delta V$ are now above the JT10 line for almost all the interval probed. For $s>0.01$ {  pc} we still
see the suggestion of a flat region, although it is now much less well defined and dispersion between bins is more prominent.

In comparing the right and left panels of Figs. 5 and 6 we see that the $\Delta V$ vs. $m_{B}$ scalings have disappeared, with $<\Delta V>$ values showing no
clear scaling with $m_{B}$. Indeed, the results of linear fits are both consistent with no dependence of $\Delta V$ on $m_{B}$, yielding $\alpha=0.089 \pm 0.21$
and $\alpha=0.086 \pm 0.49 $ with very low statistical correlations of $r=0.21$ and $r=0.12$ for the $s<0.01${  pc} and the $s>0.01${  pc} regions,
respectively. Although these power-law scalings remain consistent with the Newtonian value of 0.5 at a $2\sigma$ level, the poor correlation obtained makes
this final result more suggestive of indicating the presence of kinematic contamination, particularly when compared to the much more relevant and precise
results obtained for these scalings using the $D<125$pc sample. From Table 2 we also notice a slight increase in $<m_{B}>$ for the more distant sample,
{  showing} stars of smaller masses beginning to drop from the sample, at fixed minimum $(S/N)_{\varpi}>100$.

Thus, a combination of noisier data, and a greater probability of kinematic
contamination render wide binary samples with mean distances even $64\%$ larger than $125$pc, inadequate for the purpose of inferring fine details
regarding the scalings of their internal kinematics.

\section{Comparison to recent independent studies}

In this final section I compare the results obtained to a couple of recent studies exploring also {\it Gaia} wide binaries in the context of testing
gravity in the low acceleration regime. The first is the work of Pittordis \& Sutherland (2023), who examine velocities on the plane of the sky
for a much larger {\it Gaia} sample than the one treated here. These authors present their results in terms of distributions of the $\tilde{v}$
parameter for various $s$ cuts, where:

\begin{equation}
\tilde{v}=\frac{\Delta V}{\sqrt{G m_{B}/s}},
\end{equation}

\noindent a measure first introduced in Banik \& Zhao (2018). Clearly, $\tilde{v}=1$ for a Newtonian binary having a circular orbit {  within} the plane
of the sky. Projection effects will lead to a distribution of $\tilde{v}$ values below 1 for an observed population, while the presence of a distribution
of ellipticities will broaden the distribution, which will have a sharp upper cut at the escape velocity of $\tilde{v}=\sqrt{2}$. I now calculate
the $\tilde{v}$ values for all the 450 wide binaries appearing in the left panel of Fig. 4. The resulting sample of $\tilde{v}$ values was then divided
into two sub-samples according to the separation of the binaries, distinguishing between $s<0.01$pc and $s>0.01$pc. These two distributions are shown in
Fig. 7, with the left panel giving results for $s<0.01$ pc, and the right panel for $s>0.01$pc, red lines. The blue histograms, taken from Banik \& Zhao
(2022), give predictions for a Newtonian model and for a MOND model including the external field effect of standard AQUAL or QUMOND proposals. The numbers
in the y axis give the binned occupancy values for the wide binaries treated, in the right panel, while in the left panel, the data has been re-scaled so as to allow a
uniform comparison, by a factor of $0.66$.

For context on the external field effect (EFE) of MOND, recall that the theory is inherently non-linear, and includes an explicit acceleration scale, $a_{0}$,
with the character of dynamics depending sensitively on the ratio of the acceleration present to $a_{0}$. This implies that a small bound system immersed
within a larger mass distribution will behave in ways which are determined both by its internal acceleration, and by the local acceleration felt as part of
the larger system, beyond the inclusion of tides (see e.g. Milgrom 1986). As the details of exactly how this EFE applies are dependent on the exact particular
MOND proposal, all of which share identical predictions regarding the low acceleration behaviour of orbits about static isolated systems, e.g. Milgrom (2011)
or Milgrom (2022), it is common to find predictions for behaviour which is sensitive to the EFE presented for particular well studied proposals, such as AQUAL
or QUMOND, and also for the extreme limiting case of no EFE, e.g. Pittordis \& Sutherland (2018).

Looking at the left panel of Fig. 7 we see that the obtained distribution of $\tilde{v}$ values is broadly consistent with the Newtonian expectations, the peak
appears well within values of $\tilde{v}=1$, and no systems appear at values above $\tilde{v}=1.5$. The difference between both curves are due to the small
numbers of binaries available after the cuts imposed (Poisson noise confidence intervals on the red curve, which have not been added to avoid cluttering the
figures, make both histograms consistent), and to any offset between the assumed ellipticity distribution of the Newtonian models constructed by
Banik \& Zhao (2022) and the actual present-day $e$ distribution. The same can be said of the red and blue histograms in the right panel of Fig. 7,
which compare our results for the low acceleration $s>0.01$pc region and the AQUAL prediction of this quantity, also from Banik \& Zhao (2023). Compared to
the Newtonian $s<0.01$pc distributions, the $s>0.01$pc ones show a slight broadening, and also a slight displacement towards larger values of the $\tilde{v}$
parameters, with a {  mode} still well within $\tilde{v}=1$. Thus, our results are strongly suggestive of an AQUAL phenomenology at the low
acceleration $s>0.01$pc region.

The green histogram in this last figure, also from Banik \& Zhao (2022), shows the expectations of a MOND without external field effect model, corresponding
to the deep MOND limit of an isolated system, e.g. the models known to accurately reproduce the rotation curves of isolated spiral galaxies, e.g.
Lelli et al. (2017). With respect to the blue histograms, this last prediction shows a very distinct triangular shape, significantly shifted to the right,
with a {  mode} appearing at values of $\tilde{v}$ larger than 1, a region where the observed distribution presents almost no points. This last distribution
is clearly inconsistent, both qualitatively and quantitatively with the observed one.

The red histograms shown in Fig. 7 can now be compared to those presented recently by  Pittordis \& Sutherland (2023), their Fig. 9. We see that
results from these authors are quite similar to mine in the $s>0.01$pc right panel of Fig. 7, in the $\tilde{v}<2$ region, for all the large samples
these authors present, of about 2000 binaries, all in various sections of the low acceleration $s>0.01$pc regime. However, theirs present extended
distributions reaching values as high as $\tilde{v}=7$. This feature shows the extensive presence of kinematic contaminants in the large sample of
Pittordis \& Sutherland (2023), and alerts to their presence also within the sensitive $\tilde{v}<2$ region which has to be used to discriminate between
different gravity models. These authors are aware of this problem, and model their results through the inclusion of a fraction of hidden
tertiaries and flyby contaminants, rather than adopting the strategy followed here, of attempting the removal of these contaminants, which would necessarily
result in a much smaller sample. Thus, an statistical approach is taken where the fraction of hidden tertiaries and of flybys is allowed to vary
so that optimal values for these parameters are obtained. {  Considering three plausible ellipticity distribution functions for both Newtonian and
MOND scenarios, the final results show optimal fits with a hidden tertiary fraction of $50\%$ which are a good representation of the data for Newtonian
models. Reduced chi squared values which are within 1.2 sigma of Newtonian expectations but over 3 sigma away from MOND models for thermal ellipticity
distributions result, although the difference between the best MOND model and the worst Newtonian one is smaller than the range of values covered by
either of these two options.}

\begin{figure}
    \includegraphics[height=7.0cm,width=8.8cm]{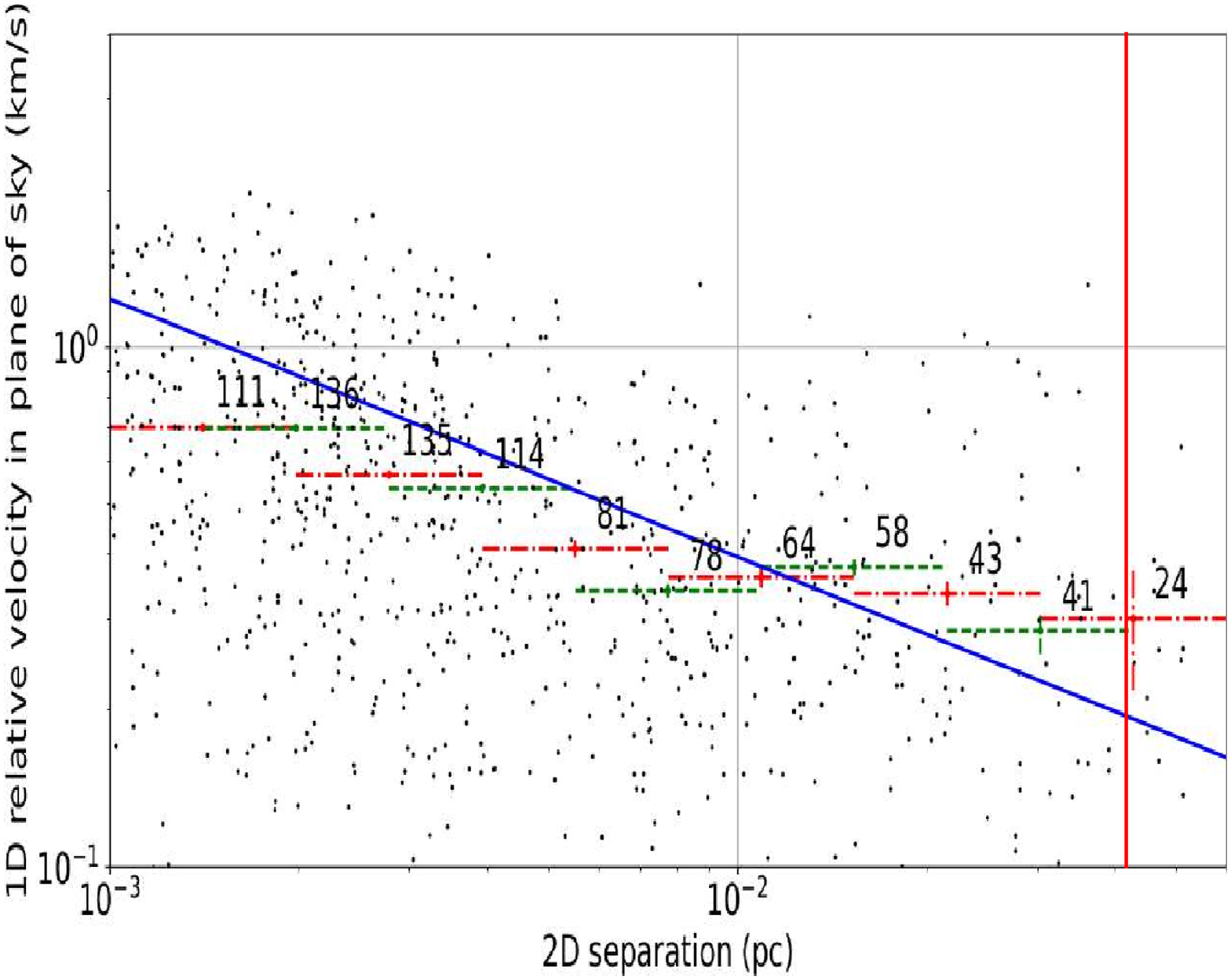}
        
    \caption{The figure shows the same data appearing in the left panel of Fig. 4, but showing this time the mean values at each bin, rather
      that the rms values of Fig. 4. For comparison, the diagonal line is the same as in Fig. 4, the vertical one again shows the MOND radius
      for the mean total mass of the binaries shown.}   
 \end{figure}

Hence, the above authors obtain results which favour standard Newtonian gravity over both MOND variants discussed above. However, having missed entirely the
Newtonian region, as the smallest $s$ value they consider {  is} $s=0.025$pc, they have no firm Newtonian point against which to calibrate or test the
robustness of the scheme used. It is quite possible that had they included a Newtonian calibration point, a regime transition at $s=0.01$pc would have
been apparent. The inclusion of a firm calibration Newtonian region in Pittordis \& Sutherland (2023) would also allow a clearer separation of the
degeneracies inherent to a problem where the fraction of hidden tertiaries, the fraction of unbound flybys and any changes in the structure of gravity
appear intermixed. {  The high best-fit hidden tertiary fraction obtained by these authors could well be a reflection of an enhanced effective
value of G in this low acceleration region. Indeed, Chae (2023) (see below), obtains a lower best-fit hidden tertiary fraction than what
Pittordis \& Sutherland (2023) find, while having a sample which extends into the high acceleration region to calibrate this number against
firm Newtonian expectations.}

Returning to Fig. 7, right panel, the clear inconsistency of the MOND without external field effect green histogram when compared to the data presented here,
shows that the analogy with a flat rotation curve galactic region suggested by the low acceleration behaviour in Fig.4, is limited. The flat behaviour
shown by the rms values of the $\Delta V$ populations in this figure is hence the result of the details of the distributions. While the deviation from
Newtonian behaviour is evident, the precise manner of this deviation has to be deduced from more than just a single moment of the distribution obtained.
The $\tilde{v}$ distributions shown in Fig. 7 make it obvious that a MOND variant including an external field effect is preferred by the data.

That the hidden tertiary cleaning strategy implemented here is highly successful, albeit leaving only a very small final
sample, can now be confirmed through three independent checks. First, the Newtonian prediction of JT10 is very accurately traced in the $s<0.01$ pc region
of the left panel of Fig. 4, in spite of the fact that the rms statistic being used in this comparison is highly
sensitive to outliers. Then, the mass-velocity power-law scaling resulting for the above mentioned sample {  is} $0.45 \pm 0.13$, in
excellent agreement with Newtonian expectations, despite the small dynamical range in mass available. Lastly, the $\tilde{v}$ distributions for all
regimes of the final binaries used show none of the extended tail at values higher than 2.6, out to 6 or even 7, which are characteristic of the results
of the Pittordis \& Sutherland group, who perform no cleaning of hidden tertiaries or flyby contaminants. In the present sample only two binaries appears
with $\tilde{v}$ values above 2, and none above 2.6. If any meaningful population of hidden tertiaries remained, none of the above three conditions would be met.

As explained in section 3, it is only through rms 1D values that the results can be compared against an existing Newtonian prediction including projection
effects, a distribution of initial ellipticities, and crucially, 10 Gyr of evolution self-consistently included in the work of JT10, who present final 1D
rms results. Note that no other wide binary analysis in the context of tests of gravity presently found in the literature has included any consideration of
the effects of dynamical evolution in the galactic environment, tidal fields and interaction with field stars, a non-negligible effect on the loosely-bound
wide binaries being studied.

The use of the 1D results of JT10 also allow a careful comparison to the high acceleration Newtonian region, as shown in the left panel of Fig. 4. This
acts as a consistency check on the study as a whole. The highly accurate agreement between present results for $s<0.01$ pc and the predictions of JT10
validates the procedure used here and shows that the final sample is highly free from kinematic contaminants of any type. This comparison is only possible
in the variables currently shown. Still, the use of the rms statistic is not entirely intuitive and lead to the mistaken interpretation in previous work
by my group of the flattening in this presentation of the data as evidence for a MOND without external field effect model. The results shown in the right
panel of Fig. 7 show this last to have been a mistaken interpretation of results.

In order to explore the character of the non-Newtonian $s>0.01$pc region better, I now repeat the left panel of Fig. 4, but showing not the rms value but the
means at each of the same binning intervals of the previous figure. This is given in Fig. 8, where the line shows again the Newtonian expectations of JT10,
for comparison. As it is the mean rather than the rms values at each bin which are being plotted, the Newtonian region now falls slightly below the rms
Newtonian expectations, but reassuringly, for $s<0.01$pc, the scaling still follows the $\Delta V \propto s^{-1/2}$ of Newtonian expectations. In going to
the low acceleration $s>0.01$pc { (2000 AU)}, just as in Fig. 4, a clear departure from Newtonian behaviour becomes apparent, although this time,
the mean $\Delta V$ values no longer remain flat, but show a fixed fractional boost over the extrapolation of the Newtonian region. In the $s>0.01$pc
region we see mean values tracing the Newtonian scaling law, but above by a small factor.

This last figure can be compared to the recent results of Chae (2023), which appeared while the present paper was being refereed. This author takes a similar
approach to Pittordis \& Sutherland (2023) discussed above, in that a large wide binary sample is maintained by not attempting a clearing of hidden tertiaries,
but in contrast to  Pittordis \& Sutherland (2023), Chae (2023) does efficiently remove flyby contamination through the use of an isolation criterion for the
binaries being considered, taken from El-Badry et al. (2021). This results in a much cleaner sample, evident from the lack of a high $\tilde{v}$ distribution,
as is also the case in my present study. The treatment of hidden tertiaries between the above two studies is similar, in that the hidden  tertiary fraction is
treated as a free parameter to be determined statistically. However, the wide binary sample considered by Chae (2023) does extend into the Newtonian regime,
reaching down to the same $s=0.001$pc lower value as I use here. This allows Chae (2023) a crucial calibration at the Newtonian region where an excellent
agreement between the data and Newtonian expectations allows an accurate calibration of the hidden tertiary fraction, as well as a consistency check of
the whole procedure.

Chae (2023) reports wide binary data accurately tracing Newtonian expectations out to $s=0.01$pc, after which a clear and highly significant departure
appears, {  which for the first time was identified in that paper as being consistent with an AQUAL MOND description of the data, as confirmed in my
present study.} Chae (2023) then presents a comparison to the first version of my present paper as it appeared on the arXiv on submission to the journal,
by plotting in his Fig. 32 his data in a $\Delta V$ vs. $s$ scatter plot, with binned values showing mean values at each bin. This last figure is qualitatively
and quantitatively consistent with the result shown in Fig. 8 here. It is very reassuring of the underlying robustness of the results obtained here that a
completely independent and highly complementary (in terms of the treatment of hidden tertiaries) study should reach conclusions consistent with those
presented here. 

We note finally that Chae (2023) intended his Fig. (31) as a comparison to Fig. 4 in the first draft of the present study, but failed to note that in
Fig. 4 of my study, for reasons already detailed, what is being plotted is the rms value at each bin and not the mean. For this reason Chae (2023)
mentions an accurate agreement with the results of the first draft of the present study in terms of an accurate tracing of the Newtonian expectations
out to $s=0.01$pc, followed by a clear and highly significant departure upwards in velocity for $s>0.01$pc { (2000 AU)}. However, Chae (2023) also
notes a qualitative difference with my earlier results, as the mean binned $\Delta V$ values do not remain flat but rather show a parallel tracing of
the Newtonian scaling, but a small factor above. This is indeed the exact behaviour shown in Fig. 8, the flat binned behaviour is found here when
plotting the rms values, not the means. Hence, agreement with Chae (2023) is complete.

\section{Discussion}

The results shown in the left panels of Figs. 4 and 5 for the $<D/pc>=90.25$ sample clearly display a gravitational anomaly in the $s>0.01$ pc region which
can not be attributed merely to the presence of noise due to random uncertainties in the {\it Gaia} data used; the mean relative velocity signal-to-noise ratio
of this sample, $<S/N>_{\Delta V}=15.7$. If random noise were determing the signal recovered, the clear Newtonian signal seen immediately before $s=0.01$
would be lost and not show the accurate tracing of the JT10 prediction right up to the appearance of the flat rms $\Delta V$ regime. Indeed, in the much noisier
$<S/N>_{\Delta V}=7.5$, $<D/pc>=147.6$ sample, we see a gradual departure upwards of the recovered $(\Delta V)_{rms}$ values which never closely trace the
JT10 prediction.

Therefore, understanding the $s>0.01$ pc scalings inferred here within a Newtonian framework requires the assumption of carefully crafted kinematic contaminants,
so as to accurately reproduce the close accordance with the scalings obtained. Pittordis \& Sutherland (2023), {  analysing only
the low acceleration $s>0.01$ pc region,} have show that the observed $\Delta V$ vs $s$ signal can be reproduced by assuming a suitable combination of unbound
flybys and hidden tertiaries as kinematic contaminants of local wide binary samples. The above authors show that an optimal fit to the non-Newtonian signal is
achieved by assuming a close to $20\%$ flyby contamination (unbound flybies as a fraction of true bound binaries) in the largest separation intervals, see their
Tables 4-6. From the left panel of Fig.4 we see that this would have to apply to binaries with internal separations below $0.06$ pc, at flyby encounter velocities
of close to $0.4$ km s$^{-1}$.

Given the mean interstellar separation of 1pc of the
Solar Neighbourhood, stars with a Gaussian velocity distribution with a $\sigma_{V}$ close to $40$ km s$^{-1}$, random encounter relative velocities will also show
a Gaussian $\Delta V$ distribution with a $\sigma_{\Delta V} = \sqrt 2 \times 40 \approx 60$ km s$^{-1}$. Hence, both in terms of spatial separations and relative
velocities, the required {  flybys} as kinematic contaminants are inconsistent with random encounters in the field. One has to assume a strong degree of correlation
invoking initial conditions, essentially a common origin for the stellar pairs making up the unbound flyby population. At current separations below $0.06$ pc and
current relative velocities of $\approx 0.4$ km s$^{-1}$, separation doubling times will be below $1.5 \times 10^{5}$ yr, a very small fraction of
$1.5 \times 10^{-5}$ the typical $10$ Gyr ages of the {  low} mass Solar Neighbourhood stars in question. It is therefore far from obvious that a fully
self-consistent Solar Neighbourhood dynamical model can be constructed where close to $10\%$ of the wide binaries presented are unbound flybies, at the observed
separations of $s<0.06$ pc and relative velocities of $\approx 0.4$ km s$^{-1}$.

Regarding hidden tertiaries as kinematic contaminants of wide binary samples ({  as originally highlighted theoretically by Banik \& Zhao 2018 and latter explicitly}
shown by Clarke 2020), Pittordis \& Sutherland (2023) have shown that a very high $50\%$ hidden tertiary percentage is required, for every bound binary formed
from two single stars, one bound binary containing a hidden tertiary is required (in addition to the amount of flybies mentioned above) to reproduce the high
velocity tail of the observed $\Delta V$ binary distribution.

Assuming such a high fraction of hidden tertiaries in our current sample is inconsistent with the estimates of Penoyre (2020) and Belokurov et al. (2020), who
estimate a hidden tertiary contamination (including even hot or outer Jupiters) below $5\%$ for a $RUWE<1.4$ cut, $D<1$kpc, using the CMD cleaning which they
propose and which we follow, in their case for the 22 month {\it Gaia} DR2 sample. Invoking a $50\%$ hidden tertiary contamination in our much cleaner 34 month
{\it Gaia} DR3 ($RUWE<1.2$, $<RUWE>=1.01$, $D<125$ pc, $<D>=90.3$ pc) sample, further restricted by the imposed internal {\it Gaia} binary probability filter of
$B_{P}<0.2$ resulting in a final $<B_{P}>=0.07$ for all the stellar sources included, from which the fastest $3\%$ $\Delta V$ values across all bins have been
removed, is highly unlikely.

\begin{table}
\begin{flushleft}
  \caption{Parameters for the mass-velocity power-law fits.}
  %  \begin{tabular}{@{} | l | llll  | @{}}
  \begin{tabular}{l l l l l}
  \hline
  \hline   
             &  $(D/pc)<125 $        &                  & $125<(D/pc)<170$       &                     \\
             &  $(s/pc)<0.01$        & $(s/pc)>0.01$         & $(s/pc)<0.01$         & $(s/pc)>0.01$            \\
            
   \hline
 $N$         & $349$            & $87$             & $321$             & $113$              \\
 $\alpha$    & $0.46\pm0.13 $ & $0.26\pm0.32$  & $0.089\pm0.21$  & $0.086\pm0.49$  \\   
  
 $r$         & $0.87$           & $0.50$           & $0.21$            & $0.21$             \\
 $<m_{B}>$   & $1.56 M_{\odot}$  & $1.56 M_{\odot}$  & $1.59 M_{\odot}$  & $1.59 M_{\odot}$     \\ 
 $\% GM_{2}$ &  $40$            &   $41$           &  $47$             &   $46$              \\
 $\% GM_{1}$ &  $61$            &   $63$           &  $67$             &   $64$              \\
   
 \hline 
 
\end{tabular} 
  The table gives the number of binaries in each of the four samples used for the fits shown in Figs. 5 and 6, together with the fitted total $<\Delta V>$ vs.
  binary mass power law index $\alpha$, the $1\sigma$ confidence interval in this quantity, and the {  Pearson} statistical correlation between $<\Delta V>$
  and total binary binned mass, $r$, average total binary mass in $M_{\odot}$ and $\%$ of binaries having both, $\% GM_{2}$, and at least one, $\% GM_{1}$,
  stellar masses from internally supplied {\it Gaia} data.

\end{flushleft}
\end{table}

If the phenomenology of the $s>0.01$ pc region is attributed to hidden tertiaries, the excellent agreement with Newtonian
predictions even immediately below $s=0.01$pc would require the abrupt termination of this hypothetical contamination, a scenario which is not supported by the
available empirical evidence which shows no relevant trend with wide binary separation on the frequency of tertiary systems, in the regimes probed to date, e.g.
Tokovinin et al. (2002), Tokovinin et al. (2010).

Finally, even if a suitable combination of kinematic contaminants can be put together to reproduce the $\Delta V$ vs. $s$ scalings seen in the left panel of
Fig. 4, any such arrangement would also have to satisfy the $\Delta V$ vs. $m_{B}$ scalings present in the data, as shown in Figs. 5 and 6.
{  Also, given the reported confidence intervals in all stellar properties as reported in {\it Gaia} DR3, which have been used in terms of a full error
propagation analysis leading to the error bars shown in all the figures presented, it appears unlikely that the signal obtained in the $s>0.01$ pc region
could be the result of error-inflated relative velocities, as $<S/N>_{\Delta V}$ values in this low acceleration region of figs. (4) and (8) are
7.31 and 7.53 for Dec. and R.A., respectively.} Thus, it is possible that we could in fact be detecting the low acceleration validity limit of standard gravity,
at a regime where the inclusion of dark matter would appear contrived. Indeed, such is the conclusion independently reached by Chae (2023), whose complementary
wide binary results closely match what has been presented here.

Following the pioneering approach of MOND (Milgrom 1983), within a classical framework, and Bekenstein (2004) in terms of covariant extensions to GR, a multitude
of modified gravity and modified inertia proposals now exist which do not require proposing a hypothetical dark matter component to explain galactic rotation
curves and cosmological observations, e.g. Moffat \& Toth (2008), Zhao \& Famaey (2010), Capozziello \& De Laurentis (2011), Verlinde (2016), Barrientos \&
Mendoza (2018), Hernandez et al. (2019b) or Skordis \& Z\l o\'{s}nik (2021), to mention but a few. Unfortunately, the available predictions of most of these,
particularly covariant extensions of GR, are limited to spherically symmetric and static metrics. Calculating modified gravity/modified inertia orbits of binary
stars within the global potential of the Milky Way is something which has only been done for a very limited set of specific MOND variants (e.g. the particular
QUMOND option explored by Banik \& Zhao 2018, {  or within the superfluid dark matter proposal which yield much the same expectations as QUMOND, as shown by
Berezhiani \& Khoury 2015}). Hence, the results presented here can not presently be compared to the majority of existing modified gravity/modified inertia proposals,
with some exceptions as presented in Pittordis \& Sutherland (2018) and in section 5 here.

Note finally that the many orders of magnitude in mass, velocity and scale which separate $a<a_{0}$ local wide binaries from the $a<a_{0}$ galactic regime
makes these interesting binaries a crucial subject for further study, which could yield surprises in terms of unknown kinematic contaminants yet to be considered
under a Newtonian framework, or important constraints to limit modified gravity scenarios helping to eventually find a complete extended theory. {  A more
careful analysis of uncertainties and of any remaining systematics in the {\it Gaia} data, as the mission timeline continues to extend and the catalogue
validation advances, will also increase the robustness of the conclusions presented, further reducing the probability of error-inflated relative velocities
or hidden tertiaries affecting our results.}

\section*{acknowledgements}

I acknowledge useful conversations with Charalambos Pittordis, Kyu-Hyun Chae, Indranil Banik, Hong-Sheng Zhao and Ricardo Cort\'{e}s on all aspects of
this work. I also acknowledge the critical input of an anonymous referee as important towards clearing up a mistaken interpretation
of my results as presented in the original version of this study, as well as having provided substantial constructive criticism leading to
a much improved final version.
All data retrieval, processing, statistical analysis and presentation was performed using software
developed jointly with Stephen Cookson. Xavier Hernandez acknowledges financial assistance from UNAM DGAPA grant IN106220 and CONACYT.
This work has made use of data from the European Space Agency (ESA) mission {\it Gaia} ({https://www.
cosmos.esa.int/gaia}), processed by the {\it Gaia} Data Processing and Analysis Consortium (DPAC, {https://www.
cosmos.esa.int/web/gaia/dpac/consortium}). Funding for the DPAC has been provided by national institutions, in
particular the institutions participating in the {\it Gaia} Multilateral Agreement.

\section*{DATA AVAILABILITY}
All data used in this work will be shared on reasonable
request to the author.

\appendix

\section{Complimentary material}

\begin{figure*}
    \includegraphics[height=7.0cm,width=8.8cm]{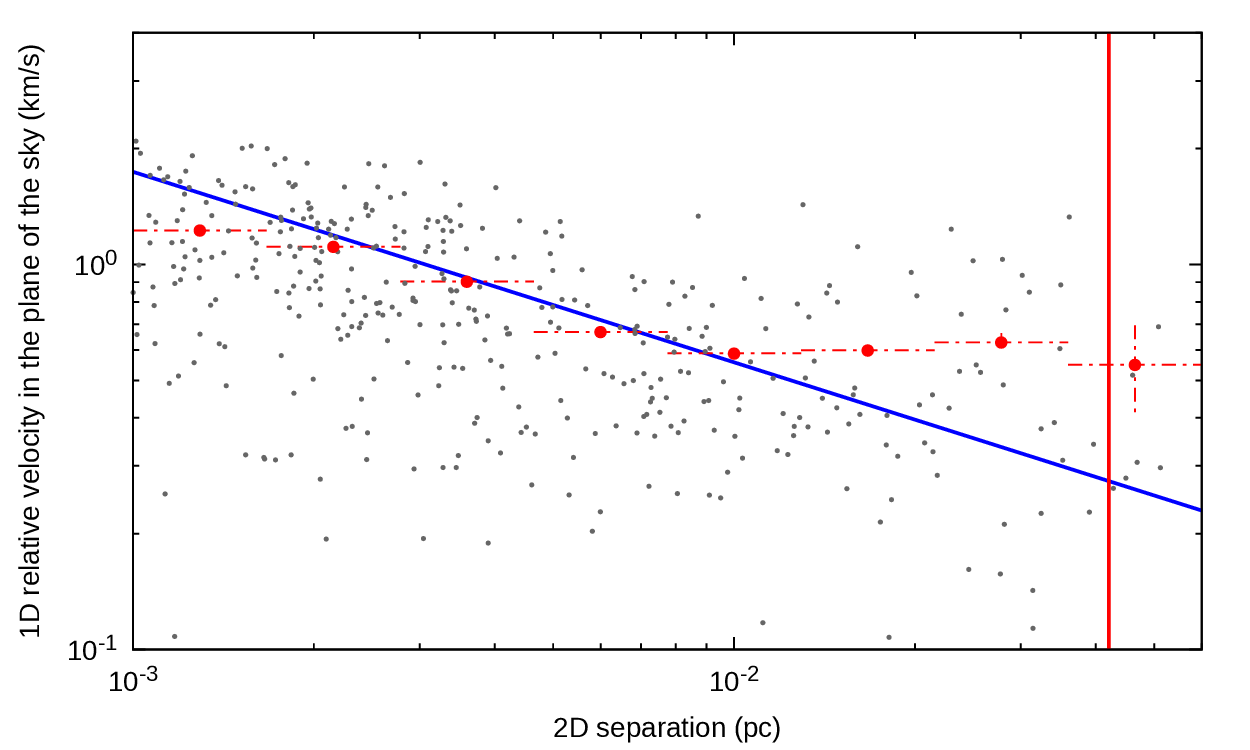}
     \hspace*{-5pt}
     \includegraphics[height=7.0cm,width=8.8cm]{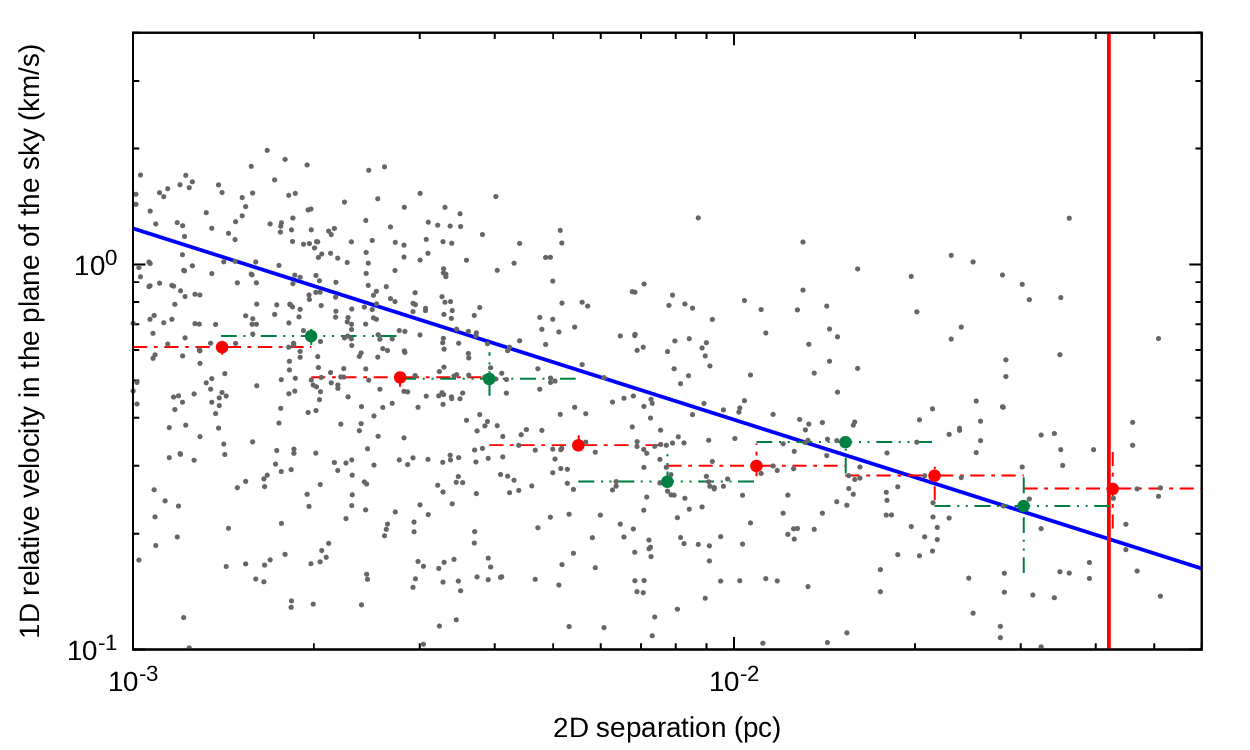}
    
     \caption{{  Left(a)}:Kinematic plot showing 2D $\Delta V$ values on the plane of the sky, dots, for binaries in the $(D/$pc$)<125$ pc sample as a
       function of the internal separation of each binary in pc, $s$, R.A. and Dec values have been added in quadrature for each binary. The points with
       error bars give the binned rms values for this same quantity, with the horizontal bars giving the bin size and the vertical ones 1$\sigma$ confidence
       intervals on the quantity being plotted. The diagonal line gives the prediction for the binned rms 2D $\Delta V$ values on the plane of the sky for
       $2M_{\odot}$ solar neighbourhood wide binaries from Jiang \& Tremaine (2010), and the vertical one shows the MOND radius for the mean total mass of
       the observed binaries shown, $<m_{B}>=1.56 M_{\odot}$. 
       {  Right(b)}:The figure shows the same data appearing in the left panel of Fig. 4, but showing this time the median values at each bin, rather
       that the rms values of Fig.(4) or the means of Fig.(8). For comparison, the diagonal line is the same as in Fig.(4), the vertical one again shows
       the MOND radius for the mean total mass of the binaries shown.}   
 \end{figure*}

{ 
To finalise, a series of variants on the results presented previously are shown here, as additional checks of the self-consistency of the
study and the results presented. First, it is clear that although R.A. and Dec. observations for a single binary represent orthogonal observations
providing two perpendicular 1D relative velocity measurements which can be used as such in terms of the projection effects of the problem, they are
not independent observations, as they refer to the same binary star. Thus, the presence of any kinematic contaminant would affect both R.A. and Dec.
measurements, and it is interesting to investigate if the signal found in the low acceleration regime is robust to the use of two data points per binary
or not. The first figure in this appendix is a repeat of Fig.(4), where R.A. and Dec. relative velocity observations for each binary have been added
in quadrature to provide a single 2D relative velocity measurement on the plane of the sky per binary. The Jiang \& Tremaine (2010) prediction has
hence been shifted upwards by a factor of $\sqrt(2)$, as there is nothing special about the R.A. Dec. directions. The reduction in the number of
data points imposes a reduction in the number of bins considered, with a corresponding reduction in separation resolution of the figure, when compared to
the original Fig. (4). Despite this, the left panel of Fig.(A1) is fully consistent with the original left panel of Fig.(4). We see the data tracing the
2D rms prediction of Jiang \& Tremaine (2010) from below in the $s<0.01$ pc (2000 AU) region, and then flattening out in the low acceleration $s>0.01$ pc
region. This validates the use of two perpendicular relative velocity entries per binary used throughout, as indeed was also the case in all the published
papers by my group on wide binaries since Hernandez et al. (2012a). In essence, each binary has been observed twice in terms of projected 1D relative
velocity on the plane of the sky, through R.A. and Dec. perpendicular velocity projections.

The next test is a remake of Fig.(8), again, using the same data as in the original, but this time showing the median value per bin rather than the means.
This is motivated by the greater robustness of rank statistics and provides a further consistency check on the robustness of our results. This is shown in
the right panel of Fig.(A1), where the diagonal line is the same Jiang \& Tremaine (2010) prediction for the Newtonian rms values as was shown in Fig.(8).
We see the medians appearing slightly below the means, as the points with error bars have shifted downwards with respect to those of Fig.(8). The $1\sigma$
confidence intervals on this rank statistic have been calculated using $j_{\pm}=(n/2)\pm (n/4)^{1/2}$, where $j_{\pm}$ are the ranks of the entries
defining the $1\sigma$ confidence intervals and $n$ is the occupancy number per bin, e.g. Conover (1999). The figure remains qualitatively the same, a
tracing of the Newtonian scaling in the $s<0.01$ (2000 AU) separation limit, and then a transition to a boosted Newtonian scaling beyond about $s=0.015$pc,
accurately tracing the AQUAL predictions, and consistent with the first mention of this feature in Chae (2023). No occupancy numbers per bin have been
given in these last two figures, as the data are the ones already shown in the left panel of Fig.(4) and in Fig.(8), and also to allow a better, uncluttered
appreciation of the trends present.

\begin{figure*}
    \includegraphics[height=7.0cm,width=8.8cm]{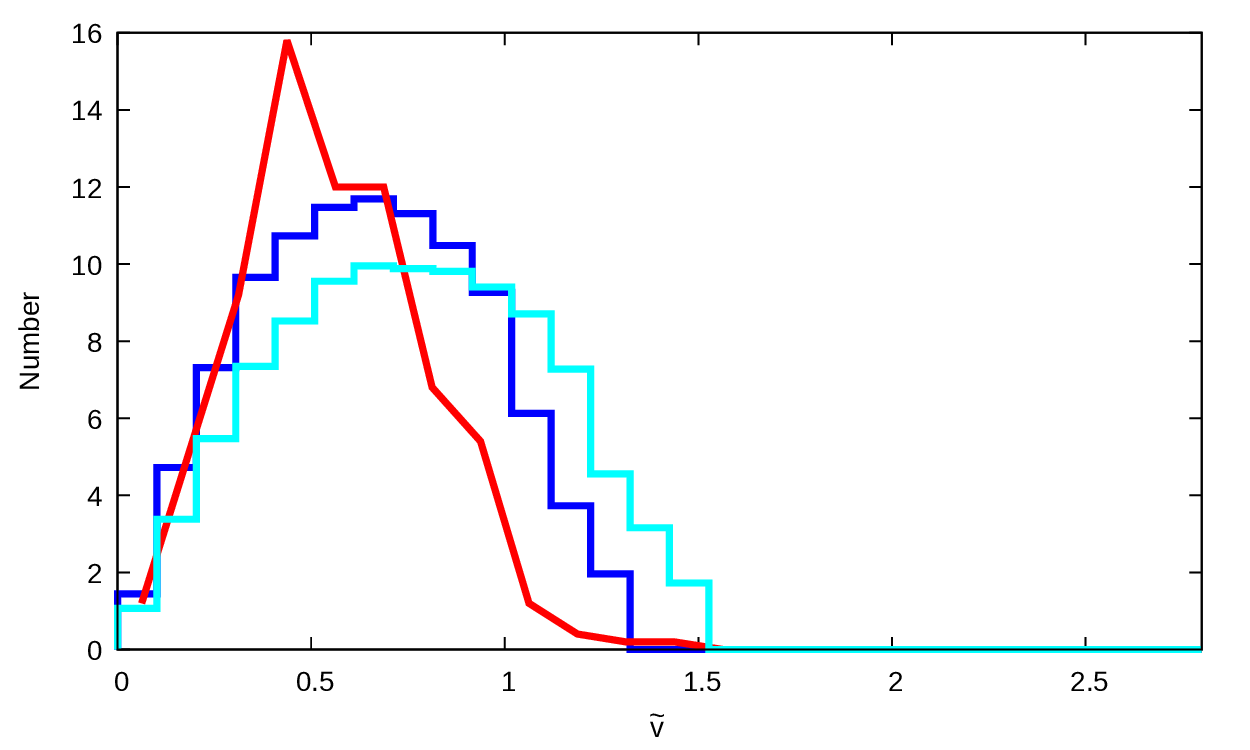}
     \hspace*{-5pt}
     \includegraphics[height=7.0cm,width=8.8cm]{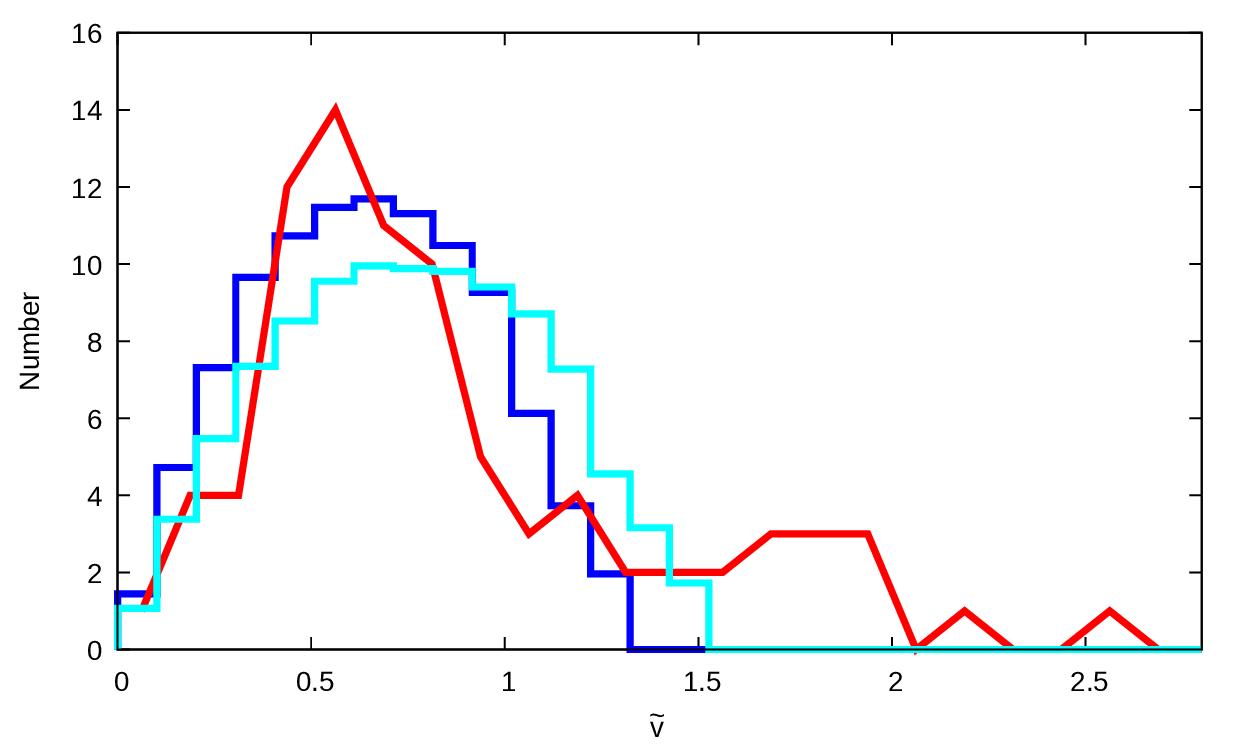}
    
     \caption{{  Left(a)}:Distribution of $\tilde{v}$ values for the wide binaries in the left panel of Fig. 4, for the $s<0.01$pc Newtonian region, red
       histogram. The blue and aqua histograms give Newtonian and MOND predictions respectively, for this distribution from Banik \& Zhao (2022). 
       {  Right(b)}::Distribution of $\tilde{v}$ values for the wide binaries in the right panel of Fig. 4, for the $s>0.01$pc low acceleration region, red
       histogram. The blue and aqua histograms give Newtonian and MOND predictions respectively, for this distribution from Banik \& Zhao (2022).}   
 \end{figure*}

The last figure in the appendix shows the $\tilde{v}$ models and observed distributions of Fig.(7), but this time both Newtonian and MOND AQUAL models are
given in each panel, with the observed distribution appearing in the left panel for the $s<0.01$ region of Fig.(4), and for the $s>0.01$ one in the right one.
The observed distributions are given by the red histograms, while the Newtonian models of Banik \& Zhao (2022) appear in blue, and the MOND AQUAL
models of these same authors in aqua. Although Newtonian and MOND predictions are quite similar in this metric, the Newtonian distribution raises faster than
the MOND one for small values of $\tilde{v}$, and indeed, the observed distribution in the $s<0.01$ pc traces this raise better than the MOND one, and better than
the observed distribution for $s>0.01$ pc, which traces better the raise of the MOND model. Similarly, in the $1.0<\tilde{v}<1.5$ region, the MOND model
predicts a larger signal than the Newtonian model, as is also seen when comparing the observed distributions for the $s>0.01$ pc region and for the
$s<0.01$pc one. Thus, we see here a closer correspondence of the MOND expectations to the observed $s>0.01pc$ region, and of the Newtonian ones to the
observations for $s<0.01$ pc data. However, it is also evident that the poor sampling afforded by the small binary candidates surviving all the rigorous
kinematic cleansing procedures implemented, makes this comparison unsuitable for drawing any conclusions. For this reason groups using this $\tilde{v}$
comparisons require binary samples running into the many thousands of stars, and for this reason too, conclusions in this present paper are drawn from
the much clearer data presentations of Figs.(4), (5), (6), (8) and (A1).

Indeed, comparing figs.(A1) and (A2) it is clear that a departure from Newtonian expectations is much easier to spot in terms of the data presentation
of Figs.(A1) than from those of Figs. (A2). Similarly as found here, the results of Chae (2023), Fig.(32), which is equivalent to Fig.(8) here,
very evidently show a departure from Newtonian expectations on reaching separations of $0.01$ pc, 2000AU, as is also the case for the very clear
acceleration plots by that same author, making a departure from Newtonian expectations for the low acceleration regime, in terms of $a_{0}$,
clearly explicit.

}

\end{document}